\documentstyle[eqsecnum,aps,epsf,epsfig,floats]{revtex} 
\begin{document}
\preprint{\vbox{\hbox{DOE/ER/40762-224}\hbox{UM PP\#01-029}}}
\title{Excited Heavy Baryons and Their Symmetries III: Phenomenology}
\author{Z.~Aziza Baccouche, Chi--Keung Chow, Thomas D.~Cohen and 
Boris A.~Gelman}
\address{Department of Physics, University of Maryland, College Park, MD 
20742-4111.} 
\maketitle
\begin{abstract}
Phenomenological applications of an effective theory of low-lying excited
states of charm and bottom isoscalar baryons are discussed at leading and 
next-to-leading order in the combined heavy quark and large $N_c$ expansion. 
The combined expansion is formulated in terms of the counting parameter
$\lambda\sim 1/m_Q, 1/N_c$; the combined expansion is in powers of 
$\lambda^{1/2}$. We work up to next-to-leading order. We obtain 
model-independent predictions for the excitation energies, the semileptonic 
form factors and electromagnetic decay rates. At leading order in the combined
expansion these observables are given in terms of one phenomenological 
constant which can be determined from the excitation energy of the first 
excited state of $\Lambda_c$ baryon. At next-to-leading order an additional 
phenomenological constant is required. The spin-averaged mass of the doublet 
of the first orbitally excited sate of $\Lambda_b$ is predicted to be 
approximately $5920 \, MeV$. It is shown that in the combined limit at leading
and next-to-leading order there is only one independent form factor describing 
$\Lambda_b \to \Lambda_c \ell \bar{\nu}$; similarly, 
$\Lambda_b \to \Lambda_{c}^{*} \ell \bar{\nu}$ and 
$\Lambda_b \to \Lambda_{c1} \ell \bar{\nu}$ decays are described by a
single independent form factor. These form factors are calculated at 
leading and next-to-leading order in the combined expansion. The value of the 
$\Lambda_b \to \Lambda_c \ell \bar{\nu}$ form factor at zero recoil is 
predicted to be $0.998$ at leading order which is very close to HQET value of 
unity. The electromagnetic decay rates of the first excited states of 
$\Lambda_c$ and $\Lambda_b$ are determined at leading and next-to leading
order. The ratio of radiative decay rates $\Gamma(\Lambda_{c}^{*} \to 
\Lambda_c \gamma) / \Gamma(\Lambda_{b1} \to \Lambda_b \gamma)$ is predicted
to be approximately $0.2$, greatly different from the heavy quark effective 
theory value of unity.
\end{abstract}

\section{Introduction and Review \label{I}}

This is a third in a series of articles in which heavy baryons (baryons 
containing a single heavy quark) are treated in the combined heavy quark and 
large $N_c$ limit \cite{hb1,hb2}. In this limit, the heavy baryon subspace of 
the QCD Hilbert space, spanned by low-lying states with baryon number one and 
heavy quark number one, exhibits an additional symmetry, viz. a contracted 
$O(8)$ symmetry \cite{hb1,hb2,hb0}. This symmetry connects orbitally excited 
states of heavy baryons to the ground state. This symmetry is distinct from 
the well-known $SU(2 N_f)$ light quark spin-flavor symmetry which connects 
heavy baryon states with light degrees of freedom in different spin and 
isopsin states, \cite{SF1,SF2,SF3,SF4,SF5}, as well as from the well-known 
$SU(2 N_h)$ heavy quark spin-flavor symmetry \cite{HQ1,HQ2,HQ3,HQ4,HQ5,HQ6}. 
An effective theory based on the contracted $O(8)$ symmetry was developed in 
Ref.~\cite{hb2}. In the real world neither the heavy quark mass, $m_Q$, nor the
number of colors, $N_c$, is infinite. However, if they are large enough, 
the heavy baryons should possess an approximate symmetry and the effective 
theory should  reliably describe heavy baryon phenomenology. It is the goal
of this paper to make predictions about spectroscopy and electroweak decays of 
heavy baryons based on the effective theory. The focus here will be on 
the phenomenology of $\Lambda_c$ and $\Lambda_b$ baryons and their excited 
states. The light degrees of freedom in these baryons are in the spin and 
isopsin ground state which simplifies the formalism. The generalization to the 
higher spin and isospin states presents no conceptual difficulties and can be 
done by combining the contracted $O(8)$ symmetry with the light quark 
spin-flavor symmetry.   

The emergence of the new symmetry can be seen in a class of models of a 
heavy baryon (which will be referred to as the bound state picture) in which
the heavy baryon is described as a bound state of a heavy meson and an ordinary
baryon \cite{bst1,bst2,bst3,bst4,bst5,bst6,bst7,bst8,bst9}. For example, the 
$\Lambda_c$ baryon is described as a bound state of a $D$ or $D^*$ meson plus
the nucleon. In the combined limit, the heavy meson and the nucleon
have large masses (the nucleon mass $m_N$ goes to infinity in the large
$N_c$ limit \cite{LN1,LN2}). The reduced mass of the meson-nucleon system is 
large, which in turn leads to a highly localized wave function. As a result, 
the potential for the low-lying excited states of the heavy baryon can 
be approximated by an harmonic potential. While the reduced mass of the system 
is of order $N_c$, the potential is of order unity \cite{LN2}. Hence, 
excitation energies are of order $N_c^{-1/2}$ and vanish as $N_c$ goes to 
infinity.
The contracted $O(8)$ group is generated by the harmonic oscillator creation 
and annihilation operators and their bilinears. In the limit where excited 
states become degenerate with the ground state, the contracted $O(8)$ group
becomes the symmetry group of the system. 

The connection between QCD and the bound state picture is obscure. However,
as was shown in Ref.~\cite{hb1}, the contracted $O(8)$ symmetry also 
emerges from QCD in the combined limit in a model-independent manner. This 
emergent symmetry along with self-consistent counting rules of Ref.~\cite{hb2}
can be used to write down an effective Hamiltonian which describes the 
low-energy excited states of heavy baryons \cite{hb2}. The effective expansion
is formulated in terms of a small parameter $\lambda$ which scales as:
\begin{equation}
\lambda \sim 1/N_c\, , \,\,\Lambda/m_Q \, ,
\label{lambda}
\end{equation}   
where $\Lambda$ is a typical hadronic scale. The actual expansion parameter
turns out to be $\lambda^{1/2}$. The ordering of heavy quark and large $N_c$ 
limits is formally irrelevant; the expansion is valid for an arbitrary value 
of $N_c \Lambda/m_Q$. 

In the combined limit the heavy baryons contain two types of 
excitations---collective and non-collective. The collective excitations can be
interpreted as the coherent motion of the brown muck (the technical term for 
the light degrees of freedom) relative to the heavy quark. In the bound state 
picture, these
excitations are described as the motion of the heavy meson relative to an
ordinary baryon. The non-collective excitations are internal excitations of 
the brown muck itself. In the bound state picture this corresponds to using an 
excited baryon in place of the nucleon ({\it e.g.} $N^*$). As shown in 
Ref.~\cite{hb2}, the typical excitation energy of the collective degrees of 
freedom near the combined limit is of order $\lambda^{1/2}$, while the typical
non-collective excitations are of order $\lambda^0$. The Hilbert space near 
the combined limit is spanned by states of the form:
\begin{equation}
|C, I \rangle = |C \rangle \otimes |I \rangle \,
\label{ST}
\end{equation}
where $C$ and $I$ stand for collective and intrinsic excitations, respectively.
The low-energy excited heavy baryons  can be identified with states
of the form of eq.~(\ref{ST}) in which only collective degrees of freedom 
are excited, {\it i.e.} $|C, 0_I \rangle$ ($0_I$ denotes the 
ground state of the brown muck). 

The effective Hamiltonian which describes
the low-energy degrees of freedom contains operators which only excite 
the collective degrees of freedom. It turns out that these operators can be 
defined in a model independent way in terms of the QCD operators near the 
combined limit \cite{hb2}. 

One group of such operators consists of the total momentum and position 
operators of the heavy baryon: $\vec P$ and $\vec X$. These operators are
conserved quantities and are well defined; they are the translation operators 
in position and momentum space. Another useful pair of operators---($\vec 
P_Q$, $\vec X_Q$)---is defined as, 
\begin{eqnarray} 
\vec{P}_Q \,& =& \, \int d^3 x \,  Q^{\dagger} (x) (-i \vec{D}) Q(x) \, ,
\nonumber \\
\vec{X}_Q \, &=& \, \int d^3 x \, Q^{\dagger}(x) \,\vec{x}\, Q(x) \, ,
\label{PQXQ}
\end{eqnarray}
where $Q(x)$ is the heavy quark field, and $\vec{D}$ is the 
three-dimensional covariant derivative. The definitions in eq.~(\ref{PQXQ})
are valid in the combined limit for the subspace of states with a heavy 
quark number one. The operators $\vec P_Q$ and $\vec X_Q$ are
(up to corrections of order $\lambda$) the translation operators for the
heavy quark in position and momentum space. The corrections arise 
because the heavy quark can not be translated independently from the rest of 
the heavy baryon. These corrections are of relative order $\lambda$ and will
be neglected in this paper, since we are working only up to relative order 
$\lambda^{1/2}$. These two pairs of operators can be used to define additional
collective variables. The operator $\vec P_{\ell}=\vec P -\vec P_Q$ and its 
canonical conjugate operator $\vec X_{\ell}$ can be interpreted as the 
momentum and position of the brown muck. This interpretation of $\vec X_{\ell}$
makes sense when the coherent motion of the brown muck is considered. The 
total position operator $\vec X$ can be expressed in terms of the operators 
$\vec X_Q$ and $\vec X_\ell$ \cite{hb1}:
\begin{equation}
\vec X = \left ({m_H \over m_N + m_H}\vec X_Q + {m_N \over m_N + m_H}
\vec X_\ell \right )(1+{\cal O}(\lambda)) \, ,
\label{X} 
\end{equation}
where $m_H$ and $m_N$ are the masses of the heavy meson and the nucleon. In 
the combined limit these masses diverge as $\lambda^{-1}$. 

There exists an additional pair of useful conjugate operators which can be 
defined in a model-independent way \cite{hb1}:
\begin{eqnarray} 
\vec{p} \, & = &\left( \frac{m_H}{m_H +m_N} \, \vec{P}  \, -  \, 
\vec{P}_Q \right ) \,\left  ( 1 \, + \, {\cal O}(\lambda) \right ) \, ,
\nonumber \\
\vec{x} \, & = &\left( \left ( 1 \, + \, \frac{m_H}{m_N} \right ) \, \
\left (\vec{X}  \, -  \vec{X}_Q \right) \right ) 
\,\left  ( 1 \, + \, {\cal O}(\lambda) \right ) \, . 
\label{xp}
\end{eqnarray} 
The operators $\vec x$ and $\vec p$ can be interpreted as the position and
momentum operators of the brown muck relative to the heavy quark. These 
operators commute with operators $\vec X$ and $\vec P$ (up to corrections of 
order $\lambda$). The operators defined in eq.~(\ref{xp}) can be re-expressed 
in terms of $\vec X_Q$, $\vec P_Q$, $\vec X_\ell$ and $\vec P_\ell$:
\begin{eqnarray}
\vec{x} & = & \vec{X}_{\ell} - \vec{X}_Q \, , 
\nonumber \\
\vec p &=&\left ({m_N \over m_N + m_H}\vec P_{\ell} - {m_H \over m_N + m_H}
\vec P_Q \right )(1+{\cal O}(\lambda)) \, .
\label{xp2}
\end{eqnarray}

While the definitions in eqs.~(\ref{X}), (\ref{xp}) and (\ref{xp2}) closely
resemble similar relations in the bound state picture, they are well-defined
(up to higher order corrections) QCD operators that excite the collective
degrees of freedom. As shown in Ref.~\cite{hb1}, the power counting
rules which lead to the effective Hamiltonian consistent with the contracted
$O(8)$ symmetry are:
\begin{eqnarray}
(x, \, X, \, X_Q, \, X_\ell) \, & \sim & \, \lambda^{1/4} \, ,
\nonumber \\
(p, \,\,  P, \,\, P_Q, \,\, P_\ell) \, & \sim & \, \lambda^{-1/4} \, .
\label{CR}
\end{eqnarray}
This means that the typical matrix elements of these operators between the
low-lying states scale as $\lambda^{1/4}$ and $\lambda^{-1/4}$ (for  
position and momentum operators, respectively). Based on these counting rules 
and the structure of the  Hilbert space in the combined limit, the most 
general effective Hamiltonian up to terms of order $\lambda$ is given by 
\cite{hb2}:  
\begin{equation}
\begin{array}{ccccccccccccccc}
 H_{\rm eff} & =  & ( m_H \, + \, m_N )& \, + \, & {c}_{0} &\,+
\, & \left (\frac{P^2}{2 (m_N + m_H)}+ \frac{p^2}{2 \mu_Q}  \, + 
\, \frac{1}{2} \kappa x^2 \right ) & \, + \, &
\frac{1}{4!} \alpha x^4 & \,  + \, & {\cal O}(\lambda^{3/2}) \, , \\  & &
\parallel & & \parallel & & \parallel & & \parallel & & \\  
 & & {\cal H}_{\lambda^{-1}} & & {\cal H}_{\lambda^{0}} 
&  & {\cal H}_{\lambda^{1/2}} & & {\cal H}_{\lambda^{1}} & &
\end{array}
\label{HEFF}
\end{equation}
where ${\cal H}_{\lambda^{n}}$ refers to the piece of the Hamiltonian whose
contribution is of order $\lambda^{n}$; the reduced mass is given by,
\begin{equation} 
\mu_Q \, = \, \frac{m_N  m_H}{m_N + m_H} \, ,
\label{mu}
\end{equation}
which is of order $\lambda^{-1}$, since $m_H$ and $m_N$ scale as 
$\lambda^{-1}$.  The coefficients $c_{0}$, $\kappa$ and $\alpha$ are of 
order $\lambda^0$. These parameters are flavor independent at order 
$\lambda^{0}$. However, they contain $1/m_Q$ corrections due to the order
$\lambda$ ambiguity both in threshold mass, $(m_N +m_H)$, and in the 
interaction of the brown muck and the heavy quark. The $1/m_Q \sim \lambda$
corrections to the constants $\kappa$ and $\alpha$ do not contribute at order
$\lambda$ as seen from eq.~(\ref{HEFF}). Hence, they can be neglected
when we work up to terms of order $\lambda$ in the effective Hamiltonian.
On the other hand, the $1/m_Q$ correction to $c_{0}$ contribute at order
$\lambda$, {\it i.e.} at the same order as the $\alpha x^4 /4!$. Since $c_{0}$
is an overall constant, the dynamics is determined by the last three terms in 
the effective Hamiltonian containing only two phenomenological 
constants---$\kappa$ and $\alpha$. 

The first term in ${\cal H}_{\lambda^{1/2}}$ describes the 
center of mass motion of the entire system and is irrelevant for the internal 
dynamics of the collective degrees of freedom. However, the counting rules 
require this kinetic term to be of order $\lambda^{1/2}$. As discussed later, 
this restriction on $\vec P$ means that the predictions of our effective 
theory for electroweak observables are valid only for velocity transfers of 
order $\lambda^{3/4}$. The terms
of order $\lambda^{1/2}$ will be referred to as leading order (LO). The 
last term is of order $\lambda$; it will be referred to as next-to-leading 
order (NLO). In this paper, we work up to NLO, {\it i.e.} we include 
corrections of relative order $\lambda^{1/2}$. The corrections
of relative order $\lambda$, which contribute only at next-to-next-to-leading 
order (NNLO) may be neglected. As can be seen from eq.~(\ref{HEFF}),
the effective expansion is an expansion in powers of $\lambda^{1/2}$
and not $\lambda$. This is ultimately connected to the intrinsic scales
that determine the dynamics: the heavy quark and the brown muck masses are of 
order $\lambda^{-1}$ while the coupling is of order unity, so that
the harmonic-like excitations are of order $\lambda^{1/2}$.

In addition to the effective Hamiltonian, a number of other important
operators (which determine the electroweak decays of heavy baryons) can be 
expressed in terms of the collective variables of eqs.~(\ref{PQXQ}), 
(\ref{X}), (\ref{xp}) and (\ref{xp2}) \cite{hb2}. In subsequent sections, we 
will apply the effective theory to study the phenomenology
of the low-lying heavy baryon states. The leading order results of our
effective theory reproduce those obtained in the bound state model of  
heavy baryons \cite{bst1,bst2,bst3,bst4,bst5,bst6,bst7,bst8,bst9}. However, 
the effective theory treatment has the advantage of being model independent.
Moreover, it can be consistently extended to higher orders. In this work, we 
extend the predictions to next-to-leading order. These corrections contribute
at relative order $\lambda^{1/2}$ and not $\lambda$ as was previously believed
\cite{bst3}.

It is important to emphasize that while $\lambda$ is formally a small number, 
in the real world $N_c=3$, so that the expansion parameter 
$\lambda^{1/2} \sim 1/\sqrt{3}$ is not particularly small. Therefore, 
predictions of the effective theory even at NLO may be relatively crude. 
Moreover, the effective theory can only be useful for describing the ground and
the first excited state of heavy baryons. The second orbitally excited state
is either unbound or very near the threshold and clearly beyond the harmonic
limit implicit in the effective theory. One also has to keep in mind that in 
the combined expansion $1/m_Q$ and $1/N_c$ corrections are formally of the 
same order. However, on phenomenological grounds, the heavy quark limit 
seems to be more closely reproduced in nature than the large $N_c$ limit. Of 
course, the ultimate test of the effectiveness of the combined expansion is
how well it reproduces experiment. At present, there is not enough data to
critically test the theory, especially in the bottom sector.  

In the next section, we discuss the spectroscopy of the low-lying states of
$\Lambda_c$ and $\Lambda_b$ baryons based on the effective Hamiltonian in 
eq.~(\ref{HEFF}). In Sec.~\ref{SD} we consider the semileptonic decays of 
these baryons . In  Sec.~\ref{RD} we further apply the effective theory to 
determine the radiative decay rates of the heavy baryons. In Sec.~\ref{C} 
we present a summary of our results. 
               
\section{Spectroscopy of Heavy Baryons  \label{S}}

The effective Hamiltonian in eq.~(\ref{HEFF}) exhibits a number of symmetries
which determine the quantum numbers and wave functions of $\Lambda_c$ and 
$\Lambda_b$ baryons and their low-energy excited states. The contracted 
$O(8)$ symmetry is generated by the set of 28 generators 
$\{ 1, a_i, a_i^\dag, T_{ij}, S_{ij}, S_{ij}^\dag \}$, where 
$T_{ij}={a_i^\dag} {a_j}$ and $S_{ij}={a_i} {a_j}$ and $i,\, j=1, 2, 3$. 
The operators $a_j$ and ${a_j}^\dag$---creation and annihilation operators of 
the three-dimensional harmonic oscillator---are defined as:
\begin{equation}
a_j = \sqrt{\mu_Q \omega_Q \over 2}\, x_j + i \sqrt{1 \over 2 \mu_Q \omega_Q} 
\,p_j \, , \qquad
a_j^\dag =\sqrt{\mu_Q \omega_Q \over 2}\, x_j - i\sqrt{1 \over 2\mu_Q \omega_Q}
\,p_j \, ,
\label{aadag}
\end{equation} 
where $\mu$, $\vec p$, $\vec x$ are defined in eqs.~(\ref{mu}) and 
(\ref{xp}); $\omega_Q =(\kappa/\mu_Q)^{1/2}$. This symmetry is broken by 
terms of order $\lambda^{1/2}$ in the effective Hamiltonian of 
eq.~(\ref{HEFF}). The $U(3)$ symmetry---a subgroup of $O(8)$ and the symmetry 
group of the three dimensional harmonic oscillator---is broken only at 
next-to-leading order. Therefore, a useful basis to compute
physical observables is an eigenbasis of the effective Hamiltonian at order
$\lambda^{1/2}$, namely the harmonic oscillator basis. This basis can be 
parameterized by the eigenvalues of the Casimir operators of the chain of the
subgroups of $U(3) \supset O(3) \supset U(1)$, {\it i.e.}
$N=T_{11}+T_{22}+T_{33}$, $L^2=L_{1}^2+L_{2}^2+L_{3}^2$ and $L_{3}$ 
(where $L_i \equiv -i\epsilon_{ijk} T_{jk}$). 
 
In the heavy quark effective theory (HQET), $\Lambda_c$ and $\Lambda_b$ 
baryons and their excited sates exhibit the $SU(4)$ heavy quark spin-flavor 
symmetry \cite{HQ1,HQ2,HQ3,HQ4,HQ5,HQ6}. This symmetry is broken by 
corrections of order  $1/m_Q$. As a result, at leading order in the pure heavy
quark expansion, the excitation energies of $\Lambda_c$ and $\Lambda_b$ 
baryons are equal. In addition, the states which differ by a heavy quark spin 
flip are degenerate. The physical origin of this symmetry is quite simple. At 
leading order in HQET, the heavy quark acts as a source of static color field 
which is independent of the heavy quark spin and flavor. The heavy quark pair 
creation and spin-dependent chromomagnetic interactions---effects that 
break spin-flavor symmetry---appear only at the next-to-leading order in HQET
({\it i.e.} at ${\cal O}(1/m_Q)$). However, in the combined heavy quark
and large $N_c$ expansion, the flavor symmetry is already broken  at leading
nontrivial order via the flavor dependence of the reduced mass $\mu_Q$. The
heavy quark is no longer a static color source but is coupled to the brown 
muck at leading order in the dynamics. However, this leading order dynamics
is still independent of the heavy quark spin. The spin-dependent 
chromomagnetic effects contribute at NNLO. Thus, the heavy quark spin, $s_Q$, 
is a good quantum number up to NNLO.
 
The eigenstates of the leading order effective Hamiltonian, eq.~(\ref{HEFF}), 
can be labeled by $|\Lambda_Q; N, l, m; \sigma \rangle$, where $N, l, m$ are 
the quantum numbers of the three-dimensional harmonic oscillator and $\sigma$ 
is the third component of the heavy quark spin ($\sigma= \pm 1/2$). The wave 
functions of these eigenstates, which will be referred to as the collective 
wave functions, are:
\begin{equation}
\Psi_{Nlm,\sigma}(\vec x) = A r^l e^{- r^2 \sqrt{\mu_Q \kappa}/ 2}
\, Y_{lm}(\theta, \phi) 
F \left (-{N-l \over 2}\, , l+{3\over 2}\, , r^2 \sqrt{\mu_Q \kappa} 
\right ) \chi (\sigma) \, ,
\label{WFls}
\end{equation} 
where $A$ is a normalization constant, $F(n, k, z)$ is a hypergeometric 
function and $\chi (\sigma)$ is a normalized spinor wave function. The spatial
part of the wave function in eq.~(\ref{WFls}) is the wave function of the
three-dimensional harmonic oscillator. There are $(N+1)(N+2)/2$ degenerate 
eigenstates for a given $N$. 

The heavy baryon states have definite total angular momentum $J$ and can be 
labeled as, $|\Lambda_Q; N, J, J_z \rangle$. These states can be written as 
linear superpositions of the states $|\Lambda_Q; N, l, m; \sigma \rangle$. The
brown muck in isoscalar charm and bottom baryons---$\Lambda_c$ and $\Lambda_b$
baryons and their excited states---has zero spin and isospin. For these 
states, the total angular momentum $J=l + s_Q=|l \pm 1/2|$. The collective 
wave functions for these states are the linear superposition of the wave 
functions in eq.~(\ref{WFls}):
\begin{eqnarray}
\Phi_{N, l+1/2, J_z}& = &\sqrt{J+J_z\over 2 J_z}\Psi_{N, J_{z}-1/2, \sigma=1/2}
+\sqrt{J-J_z\over 2 J_z}\Psi_{N, J_{z}+1/2, \sigma=-1/2} \, ,
\nonumber \\
\Phi_{N, l-1/2, J_z}& = &\sqrt{J-J_z+1 \over 2 J_z +2}\Psi_{N, J_{z}-1/2, 
\sigma=1/2}
+\sqrt{J+J_z+1\over 2 J_z + 2}\Psi_{N, J_{z}+1/2, \sigma=-1/2} \, ,
\label{WF}
\end{eqnarray}
where the square factors are the appropriate Clebsch-Gordan coefficients.

In the charm sector, the states 
$|\Lambda_c; 0, {1 \over 2}, J_z \rangle$,  $|\Lambda_c; 1, {1 \over 2}, 
J_z \rangle$ and $|\Lambda_c; 1, {3 \over 2}, J_z \rangle$ correspond to the 
experimentally observed states $\Lambda_c$ ($J^P={1 \over 2}^{+}$), 
${\Lambda_c}(2593)$ ($J^P={1 \over 2}^{-}$) and ${\Lambda_c}(2625)$ 
($J^P={3 \over 2}^{-}$), respectively \cite{exp1,exp2,exp3}. The states 
$|\Lambda_c; 0, {1 \over 2}, J_z \rangle$ and $|\Lambda_c; 1, {1 \over 2}, 
J_z \rangle$ are degenerate in the combined heavy quark and large $N_c$ limit 
due to the heavy quark spin symmetry valid up to NNLO. The same 
parameterization of states exist in the bottom sector; however only the ground
state, $\Lambda_b$ ($J^P={1 \over 2}^{+}$), has been observed to date. The lack
of the observation of the excited states limits our ability to test the theory
at present. For further convenience a shorthand notation for heavy baryon 
states will be used. The ground state will be denoted as,
\begin{equation}
|\Lambda_Q \rangle \equiv |\Lambda_Q; 0, {1 \over 2}, J_z \rangle \, ,
\label{groundstate}
\end{equation}
and the doublet of the first excited state will be denoted as,
\begin{eqnarray}
|\Lambda_{Q1} \rangle & \equiv &|\Lambda_Q; 1, {1 \over 2}, J_z \rangle \,,
\nonumber \\
|\Lambda_{Q1}^{*} \rangle & \equiv &|\Lambda_Q; 1, {3 \over 2}, J_z \rangle \,.
\label{doublet}
\end{eqnarray}

The expansion of the effective Hamiltonian, eq.~(\ref{HEFF}), including terms 
of order $\lambda$ contains three phenomenological parameters---$c_{0}$,
$\kappa$, $\alpha$. The first two terms in eq.~(\ref{HEFF}) 
determine the dissociation threshold of the heavy baryon into a heavy meson 
and an ordinary baryon. There are two heavy mesons for each heavy quark 
flavor---pseudoscalar and pseudovector mesons. In the charm sector they
correspond to $D$ and $D^*$ mesons and in the bottom sector to $B$ and $B^*$. 
The pseudoscalar and pseudovector mesons differ due to the heavy quark spin 
flip. In the combined limit these mesons are degenerate up to 
corrections of relative order $\lambda$. This leads to an ambiguity in the 
energy of the dissociation threshold. One natural way to define
this threshold is to use the spin-averaged mass of the meson doublet:
\begin{eqnarray} 
m_{\bar{D}} & \equiv & {1\over 4} m_D + {3\over 4} m_{D^*} \approx 1980 \, MeV
\, , \nonumber \\
m_{\bar{B}} & \equiv & {1\over 4} m_B + {3\over 4} m_{B^*} \approx 5310 \, MeV
\, ,
\label{SAM}
\end{eqnarray}
where the masses are averaged over one spin state of a pseudoscalar
meson and three states of the pseudovector meson. Similarly, the spin-averaged
mass of the doublet of the first orbitally excited states 
$\Lambda_{Q1}$ and $\Lambda_{Q1}^{*}$ is given by,
\begin{equation} 
m_{\bar{\Lambda}_{Q}^{*}}  \equiv {1\over 3} m_{\Lambda_Q^{*}} + 
{2\over 3} m_{\Lambda_{Q1}^{*}} \, ,
\label{BSAM}
\end{equation}
where the mass is averaged over the two spin states with $J^{P}={1\over2}^-$ 
and four possible states with $J^{P}={3\over2}^-$. The spin-averaged mass of 
the $\Lambda_{c1}$ and $\Lambda_{c1}^{*}$ is:
\begin{equation} 
m_{\bar{\Lambda}_{c}^{*}} = {1\over 3} m_{\Lambda_{c1}} + 
{2\over 3} m_{\Lambda_{c1}^{*}} \approx 2610 \, MeV \, .
\label{lambdacSAM}
\end{equation}

The ground state mass and the spin-averaged mass of the first excited state 
can be determined from the effective Hamiltonian in eq.~(\ref{HEFF}) including
terms of order $\lambda$:
\begin{eqnarray}
m_{\Lambda_Q}& = & m_N+m_H+c_{0}+{3 \over 2}\sqrt{\kappa \over\mu_Q}+
{15\over 96}{\alpha \over \kappa \mu_Q}+{\cal O}(\lambda^{3/2}) \, ,
\nonumber \\
m_{\bar{\Lambda}_{Q}^{*}} & = & m_N+m_H+c_{0}+
{5 \over 2}\sqrt{\kappa \over \mu_Q}+
{35\over 96} {\alpha \over \kappa \mu_Q}+{\cal O}(\lambda^{3/2}) \, ,
\label{MGME}
\end{eqnarray}
where the mass $m_H$ is the spin-averaged mass of the meson doublet, 
eq.~(\ref{SAM}). The ${\cal O} (\lambda^{1/2})$ term in eq.~(\ref{MGME}) 
corresponds to the energy of the ground and the first excited states of the 
three dimensional harmonic oscillator. The correction of order $\lambda$ is 
obtained by treating the term $\alpha x^4 /4!$ perturbatively. 

The expressions for the heavy baryon masses in eq.~(\ref{MGME})
lead to a number of predictions at each order in the combined 
expansion. At order $\lambda^0$, {\it i.e.} neglecting all nontrivial terms,
we have the following relation:
\begin{equation}
(m_{\Lambda_b} - m_{\Lambda_c}) - (m_{\bar{B}} - m_{\bar{D}})={\cal O}
(\lambda^{1/2}).
\label{P1}
\end{equation}
This is a well-known result in HQET which is a consequence of the heavy 
quark flavor symmetry. In the combined limit, the flavor symmetry is already 
broken at order $\lambda^{1/2}$. It is important to emphasize that the 
prediction in eq.~(\ref{P1}) is a statement about the flavor independence
of the constant $c_{0}$ (intuitively the binding energy of the ground state 
of the heavy baryon at ${\cal O}(\lambda^0)$) at order $\lambda^0$. 
Equation~(\ref{P1}) is well satisfied in nature: 
$(m_{\Lambda_b} - m_{\Lambda_c}) - (m_{\bar{B}} - m_{\bar{D}})=10 \, MeV$. 
This should be compared to the mass difference of the heavy meson doublet 
which is $140 \,  MeV$ in the charm sector and $40 \, MeV$ in the bottom 
sector. This mass difference is of relative order $\lambda$.

At order $\lambda^{1/2}$---the leading nontrivial order---there are 
additional relations between the spectroscopic observables. One of them is a 
relation between the excitation energies of $\Lambda_c$ and $\Lambda_b$ 
baryons. Equations~(\ref{MGME}) imply,
\begin{equation}
m_{\bar{\Lambda}_{b}^{*}} - m_{\Lambda_b} =(m_{\bar{\Lambda}_{c}^{*}} - 
m_{\Lambda_c}) \sqrt{\mu_c \over \mu_b}\left(1 + {\cal O}(\lambda)\right) \, .
\label{P2}
\end{equation}   
In HQET the excitation energies of $\Lambda_c$ and $\Lambda_b$ baryons are 
equal up to $1/m_Q$ corrections. Equation~(\ref{P2}) predicts (up to NLO) 
the first excited state of $\Lambda_b$ (or more precisely, the spin-averaged 
mass) to lie approximately $300 \, MeV$ above the ground state. Hence, the 
model independent prediction at LO for the spin-averaged mass of the first
orbitally excited state of $\Lambda_b$ is:
\begin{equation}
m_{\bar{\Lambda}_{b}^{*}}\approx 5920 \, MeV 
\left (1 + {\cal O}(\lambda)\right) \, .
\label{massLbstar}
\end{equation}   

Other relations can be obtained at this order by combining 
eq.(\ref{MGME}) for two heavy quark flavors and re-expressing the results in 
terms of physical observables:
\begin{eqnarray}
(m_{\Lambda_b} - m_{\Lambda_c}) - (m_{\bar{B}} - m_{\bar{D}}) & - & 
{3 \over 2} (m_{\bar{\Lambda}_{c}^{*}} - m_{\Lambda_c})(\sqrt{\mu_c \over 
\mu_b}-1)= {\cal O}(\lambda) \, ,
\nonumber \\
(m_{\bar{\Lambda}_{b}^{*}} - m_{\bar{\Lambda}_{c}^{*}}) - (m_{\bar{B}}- 
m_{\bar{D}}) & - & 
{5 \over 2} (m_{\bar{\Lambda}_{c}^{*}} - m_{\Lambda_c})(\sqrt{\mu_c \over 
\mu_b}-1)= {\cal O}(\lambda) \, ,
\label{P3}
\end{eqnarray}
which are the NLO corrections to eq.~(\ref{P1}). The first of the 
relations in eq.~(\ref{P3}) shows that the leading order prediction of HQET 
(eq.~(\ref{P1})) in the combined limit is 
already broken at order $\lambda^{1/2}$. It is interesting to note that
this relation is satisfied with a bigger error than that in eq.~(\ref{P1}).
Indeed, $(m_{\Lambda_b} - m_{\Lambda_c}) - (m_{\bar{B}} - m_{\bar{D}})-
(3 / 2) (m_{\bar{\Lambda}_{c}^{*}} - m_{\Lambda_c})((\mu_c /\mu_b)^{1/2}-1)= 
60 \, MeV$. However, this is still consistent with zero within the theoretical
uncertainty which can be as high as $140 \, MeV$.

At order $\lambda$ there is one additional parameter in the effective 
Hamiltonian---the constant $\alpha$. The excitation energies in the 
charm and bottom sectors at this order are:
\begin{eqnarray}
m_{\bar{\Lambda}_{c}^{*}} - m_{\Lambda_c} & = &
\sqrt{\kappa \over \mu_c} + {5 \over 4!}{\alpha \over \kappa \mu_c}+
{\cal O}(\lambda^{3/2}) \, ,
\nonumber \\
m_{\bar{\Lambda}_{b}^{*}} - m_{\Lambda_b} & = &
\sqrt{\kappa \over \mu_b} + {5 \over 4!}{\alpha \over \kappa \mu_b}+
{\cal O}(\lambda^{3/2}) \,.
\label{P4}
\end{eqnarray}
The excitation energies of $\Lambda_c$ and $\Lambda_b$ completely determine 
two phenomenological parameters $\kappa$ and $\alpha$. These constants can 
then be used to predict the semileptonic form factors and radiative
decay rates which will be discussed in the following sections.

\section{Semileptonic decays \label{SD}}

\subsection{Decays of the Ground State Heavy Baryons \label{SDGG}}

The semileptonic decays of heavy baryons were extensively analyzed within 
the framework of HQET \cite{HQ3,HQ4,HQ5,HQ6}. The heavy quark spin-flavor 
symmetry imposes very strict constraints on the number of independent 
semileptonic form factors, 
on their normalization and their functional dependence. For example, there is 
only one form factor for the $\Lambda_b \to \Lambda_c \ell \bar{\nu}$ decay at 
leading order in the $1/m_Q$ expansion. This form factor is a smooth function 
of the recoil parameter $w = v \cdot v^\prime$, where $v$ 
and $v^\prime$ are the $4$-velocities of the initial and final baryons. Indeed,
as $m_Q \to \infty$, the effect of the heavy quark current on the heavy baryon
is a boost of the heavy quark from an initial velocity $v$ to a final 
velocity $v^\prime$. Due to the heavy quark spin-flavor symmetry, the 
transition matrix element is independent of the heavy quark spin and flavor. 
In the limit as $m_Q \to \infty$, the matrix element is determined by 
the overlap of the final and initial states of the brown muck, which depends
only on the heavy quark velocity. Thus, the form factors can depend only on 
the recoil parameter, $w$. Moreover, at zero recoil ($w =1$) the brown muck 
does not feel any change---it is insensitive to the heavy quark spin 
and flavor---so that the transition matrix element (with properly normalized 
initial and final states) is unity. This gives a unique nonperturbative 
normalization to the form factors \cite{HQ3,norm}. By Luke's theorem,
this normalization is unchanged at next-to-leading order in the heavy 
quark expansion \cite{mQ1}. From what follows, it will become
clear that the situation is quite different in the combined limit.

Let us first consider the semileptonic decay of the ground state of the
$\Lambda_b$ to the ground state of $\Lambda_c$ in which the initial and final
baryons have 4-velocities  $v$ and $v^\prime$.
The hadronic part of the invariant amplitude of such a transition
is determined by the matrix element of the left-handed current 
$J=\bar c \gamma^\mu (1-\gamma_5) b$. This matrix element is conventionally 
parameterized by six form factors:
\begin{equation}
\langle \Lambda_c (\vec{v}^\prime)|\bar c \gamma^\mu (1-\gamma_5) b |
\Lambda_b (\vec{v}) \rangle = 
\bar{u}_c (\vec{v}^\prime)\left (\Gamma_V - \Gamma_A \right ) u_b (\vec{v})\,, 
\label{JGVGA} 
\end{equation}
where 
\begin{eqnarray}
\Gamma_V & = & f_1 \gamma^\mu +i f_2 \sigma^{\mu\nu} q_\nu + f_3 q^\mu \, ,
\nonumber \\
\Gamma_A & = & \left ( g_1 \gamma^\mu +i g_2 \sigma^{\mu\nu} q_\nu +
g_3 q^\mu \right )\gamma_5  \, , 
\label{J} 
\end{eqnarray}
where $q=m_{\Lambda_c} v^\prime -m_{\Lambda_b} v$ is the momentum transfer and
$u_b (\vec{v}^\prime)$, $u_c(\vec{v})$ are the Dirac spinors corresponding to 
$\Lambda_c$ and $\Lambda_b$ baryons. The spinors are normalized as,
\begin{equation}
u^{\dag}_{Q} (\vec{v}, s^\prime ) u_{Q}(\vec{v}, s)= {E_p \over M_Q} 
\delta^{s^\prime s} \, .
\label{spnorm}
\end{equation}

In HQET, the form factors (as functions of the velocity transfer) have a smooth
expansion in powers of $1/m_Q$. As will be seen from the 
analytic expressions, this is no longer true for the $\lambda$-expansion 
in the combined limit. In fact, the form factors as functions of the velocity 
transfer have an essential singularity in the combined limit. 
Intuitively, this can be seen by considering the semileptonic decay in the 
bound state picture. The amplitude is an overlap of the harmonic oscillator 
wave functions. The final wave function contains a factor 
$e^{-i m_N {\vec x}\cdot {\vec v}}$ due to the boost operator. As a result, 
the amplitude in the combined limit is proportional to  
\begin{equation}
\exp\left ({-{m_{N}^{2} |\delta \vec v|^2 \over 2(\sqrt{\kappa \mu_b}+
\sqrt{\kappa \mu_c})}}\right) \sim 
\exp \left ( {- \lambda^{-3/2} |\delta \vec v|^2 } \right ) \, ,
\label{exp}
\end{equation}
where $|\delta \vec v| \equiv |\vec{v}^\prime - \vec{v}|$ is the velocity 
transfer. 
Thus, the amplitude is exponentially small for velocity transfers of order 
unity and can not be reliably determined using an expansion in powers of 
$\lambda^{1/2}$. On the other hand, for velocity transfers of order 
$\lambda^{3/4}$ the semileptonic matrix is of order unity \cite{bst3}.

There is another reason to consider only velocity transfers of order 
$\lambda^{3/4}$. The effective Hamiltonian in eq.~(\ref{HEFF}) is 
based on the self-consistent counting rules of eq.~(\ref{CR}) according to 
which all the momentum operators (including $\vec P$) are of order 
$\lambda^{-1/4}$. As was discussed in Sec.~\ref{I}, the kinetic term 
corresponding to the center-of-mass motion is of order $\lambda^{1/2}$. Thus, 
the counting rules require a typical heavy baryon velocity to be of order 
$\lambda^{3/4}$ for the expansion to be valid. In what follows, we will only
consider $|\delta \vec v|$ of order $\lambda^{3/4}$.

In HQET, the heavy quark spin symmetry is used to reduce the six form factors
in eq.~(\ref{J}) to a single independent Isgur-Wise function with corrections 
of order $1/m_Q$ \cite{HQ3,HQ4,HQ5,HQ6}. In the combined limit, the heavy 
quark 
symmetry generates corrections of order $\lambda$, while, as was discussed 
above, the hadronic matrix element as a function of the velocity transfer is 
exponentially suppressed. Nevertheless, as will be shown shortly, for the 
velocities of order $\lambda^{3/4}$, ({\it i.e} in the regime of the 
applicability of the combined expansion) one form factor determines the 
hadronic amplitude in eq.~(\ref{JGVGA}); moreover, this form factor is 
calculable in a model-independent way near the combined limit. As will become 
clear, this is a 
consequence of the self-consistent scaling of the Lorentz components of the 
amplitude. In this velocity regime, it is more convenient to use a different
parameterization of the matrix element of left-handed current, 
eq.~(\ref{J}). The form factors can be re-expressed using two Dirac-like 
equations,
\begin{eqnarray}
{\bar u}_c (\vec{v}^\prime) \left (i \sigma^{\mu\nu} p_\nu +q^\mu + 
\gamma^\mu (m_{\Lambda_c} - m_{\Lambda_b}) \right ) u_b (\vec{v})& = & 0 \, ,
\nonumber \\
{\bar u}_c (\vec{v}^\prime) \left (i \sigma^{\mu\nu} q_\nu +p^\mu + 
\gamma^\mu (m_{\Lambda_b} - m_{\Lambda_c}) \right )
\gamma_5 u_b (\vec{v}) & = & 0 \, ,
\label{Dirac}
\end{eqnarray}
where $p=m_{\Lambda_c} v^\prime + m_{\Lambda_b} v$. This 4-vector $p$ should
not be confused with the collective variable $\vec p$ introduced in 
Sec.~\ref{I}. Using these equations, 
the vector and axial currents in eq.~(\ref{J}) can be written as,
\begin{eqnarray}
\Gamma_V & = & F_1 \gamma^\mu +i F_2 \sigma^{\mu\nu} q_\nu +
i F_3  \sigma^{\mu\nu} p_\nu \, ,
\nonumber \\ 
\Gamma_A &= &\left(G_1 \gamma^\mu + G_2 p^\mu + G_3 q^\mu \right )\gamma_5\, ,
\label{JNR} 
\end{eqnarray}
where the form factors $f_i$ and $g_i$ are expressed in terms of $F_i$ and 
$G_i$ by,
\begin{eqnarray}
f_1=F_1+F_3(m_{\Lambda_c} -m_{\Lambda_b}) , \,\,\,\,\,\, f_2=F_2, \,\,\,\,\,\,
f_3=-F_3  ,
\nonumber \\
g_1=G_1+G_2(m_{\Lambda_c} - m_{\Lambda_b}),\,\,\,\,\,\, g_2=-G_2, \,\,\,\,\,\, 
g_3=G_3  .
\label{FGfg}
\end{eqnarray}
Furthermore, it is useful to consider separately the time and space components
of the vector and axial currents separately:
\begin{equation}
\langle \Lambda_c (\vec{v}^\prime)|\bar{c}\gamma^0 b|\Lambda_b (\vec{v}) 
\rangle =
\bar{u}_c (\vec{v}^\prime) \left (F_1 \gamma^0 + F_2 \alpha^j q^j + F_3  
\alpha^j p^j \right ) u_b (\vec{v}) \, ,
\label{JV0} 
\end{equation}
\begin{equation}
\langle \Lambda_c (\vec{v}^\prime)|\bar{c}\gamma^i b|
\Lambda_b (\vec{v}) \rangle =
\bar{u}_c (\vec{v}^\prime) \left (F_1 \gamma^i +F_2 \left(\alpha^i q^0+
\epsilon^{ijk} \Sigma^j q^k \right )
+ F_3  \left(\alpha^i p^0+\epsilon^{ijk}\Sigma^j p^k \right) \right) u_b 
(\vec{v})\, ,
\label{JVV} 
\end{equation}  
\begin{equation}
\langle \Lambda_c (\vec{v}^\prime)|\bar{c}\gamma^0 \gamma_5 b|\Lambda_b 
(\vec{v})\rangle
= \bar{u}_c (\vec{v}^\prime)\left (G_1 \gamma^0 + G_2 p^0 + G_3 q^0 \right )
\gamma_5 u_b (\vec{v}) \, ,
\label{JA0} 
\end{equation} 
\begin{equation}
\langle \Lambda_c (\vec{v}^\prime)|\bar{c}\gamma^i\gamma_5 b|\Lambda_b 
(\vec{v}) \rangle =
\bar{u}_c (\vec{v}^\prime)\left (G_1 \gamma^i + G_2 p^i + G_3 q^i  \right ) 
\gamma_5 u_b (\vec{v}) \, ,
\label{JAV} 
\end{equation}   
where $\alpha^i=\gamma^0\gamma^i$ and $\Sigma^k=i \epsilon^{ijk}\sigma^{ij}
=\alpha^k \gamma_5$.

Out of the six form factors in eqs.~(\ref{JV0}), (\ref{JVV}), (\ref{JA0}) and 
(\ref{JAV}), only two, namely $F_1$ and $G_1$, contribute at leading order in 
the combined expansion if the velocity transfer is of order $\lambda^{3/4}$. 
Indeed, as shown in the Appendix, the matrix elements satisfy the following 
scaling rules in the combined limit:
\begin{eqnarray}
\langle \Lambda_c (\vec{v}^\prime)|\bar{c}\gamma^0 b|\Lambda_b (\vec{v}) 
\rangle & \sim & \lambda^{0}\, ,
\nonumber \\
\langle \Lambda_c (\vec{v}^\prime)|\bar{c}\gamma^i b|
\Lambda_b (\vec{v}) \rangle & \sim & \lambda^{3/4} \, ,
\nonumber \\
\langle \Lambda_c (\vec{v}^\prime)|\bar{c}\gamma^0 \gamma_5 b|\Lambda_b 
(\vec{v})\rangle & \sim & \lambda^{3/4} \, ,
\nonumber \\
\langle \Lambda_c (\vec{v}^\prime)|\bar{c}\gamma^i\gamma_5 b|\Lambda_b 
(\vec{v}) \rangle  & \sim & \lambda^0 .
\label{CRLHS}
\end{eqnarray}   

These rules lead to a particular scaling of the form factors which can be 
determined by requiring a consistent scaling of the left- and right-hand side 
of eqs.~(\ref{JV0}), (\ref{JVV}), (\ref{JA0}) and (\ref{JAV}). The invariant 
kinematic factors containing Dirac spinors and Dirac matrices scale as:
\begin{eqnarray}
\bar{u}_c (\vec{v}^\prime)\gamma^0 u_b (\vec{v}) & \sim &
\lambda^{0},\,\,\,\,\,\,\,\,\,\,\,\,\,\,\,\,\,\,\,\,\,\,\,\,\,\,\,
\,\,\,\,\,\,\,\,\,\,\,\,\,\,\,\,
\bar{u}_c (\vec{v}^\prime)\gamma^0 \gamma_5 u_b (\vec{v})  \sim \lambda^{3/4} 
\, ,
\nonumber \\
\bar{u}_c (\vec{v}^\prime)\alpha^j q^j u_b (\vec{v}) & \sim &
\lambda^{1/2},
\,\,\,\,\,\,\,\,\,\,\,\,\,\,\,\,\,\,\,\,\,\,\,\,\,\,\,
\,\,\,\,\,\,\,\,\,\,\,\,
\bar{u}_c (\vec{v}^\prime) p^0 \gamma_5 u_b (\vec{v}) 
\sim \lambda^{-1/4} \, ,
\nonumber \\
\bar{u}_c (\vec{v}^\prime)\alpha^j p^j u_b (\vec{v}) & \sim &
\lambda^{1/2},
\,\,\,\,\,\,\,\,\,\,\,\,\,\,\,\,\,\,\,\,\,\,\,\,\,\,\,
\,\,\,\,\,\,\,\,\,\,\,\,
\bar{u}_c (\vec{v}^\prime) q^0 \gamma_5 u_b (\vec{v}) 
\sim \lambda^{-1/4} \, ,
\nonumber \\
\bar{u}_c (\vec{v}^\prime)\gamma^i u_b (\vec{v}) & \sim & \lambda^{3/4},
\,\,\,\,\,\,\,\,\,\,\,\,\,\,\,\,\,\,\,\,\,\,\,\,\,\,\,
\,\,\,\,\,\,\,\,\,\,\,\,
\bar{u}_c (\vec{v}^\prime)\gamma^i \gamma_5 u_b (\vec{v})   \sim  \lambda^{0} 
\,,
\nonumber \\
\bar{u}_c (\vec{v}^\prime)\left(\alpha^i q^0+\epsilon^{ijk} \Sigma^j q^k 
\right ) u_b (\vec{v})
& \sim &  \lambda^{-1/4},  
\,\,\,\,\,\,\,\,\,\,\,\,\,\,\,\,\,\,\,\,\,\,\,\,\,\,\,
\,\,\,\,\,\,\,\,\,\,
\bar{u}_c (\vec{v}^\prime) p^i \gamma_5 u_b (\vec{v}) \sim  \lambda^{1/2} \, 
\nonumber \\
\bar{u}_c (\vec{v}^\prime)\left(\alpha^i p^0+\epsilon^{ijk} \Sigma^j p^k 
\right ) u_b (\vec{v})
& \sim &  \lambda^{-1/4},  
\,\,\,\,\,\,\,\,\,\,\,\,\,\,\,\,\,\,\,\,\,\,\,\,\,\,\,
\,\,\,\,\,\,\,\,\,\,
\bar{u}_c (\vec{v}^\prime) q^i \gamma_5 u_b (\vec{v}) \sim  \lambda^{1/2} \, .
\label{KCR}
\end{eqnarray}
Combining the scaling rules of eq.~(\ref{KCR}) with the scaling of the matrix 
elements in eqs.~(\ref{JVV}) and (\ref{JA0}), we obtain:
\begin{eqnarray}
F_1 & \sim & \lambda^0 , \,\,\,\,\, F_2 \sim \lambda , \,\,\,\,\,
 F_3 \sim \lambda \, ,  
\nonumber \\
G_1 & \sim & \lambda^0 , \,\,\,\,\,  G_2 \sim  \lambda , \,\,\,\,\,
G_3  \sim \lambda \, .
\label{FFCR}
\end{eqnarray}
It is easy to see that these scaling rules are consistent with scaling of the
amplitudes in eqs.~(\ref{JV0}) and (\ref{JAV}).

Using the $\lambda$-scaling of the form factors in eq.~(\ref{FFCR}), the only 
dominant matrix elements at NLO are given by,
\begin{eqnarray}
\langle \Lambda_c (\vec{v}^\prime)|\bar{c}\gamma^0 b|\Lambda_b (\vec{v}) 
\rangle &=&
F_1 \bar{u}_c (\vec{v}^\prime) \gamma^0 u_b (\vec{v})
\left( 1 +{\cal O}(\lambda^{3/2})\right)\, ,
\nonumber \\
\langle \Lambda_c (\vec{v}^\prime)|\bar{c}\gamma^i \gamma_5 b|\Lambda_b 
(\vec{v}) 
\rangle &=&
G_1 \bar{u}_c (\vec{v}^\prime) \gamma^i \gamma_5 u_b (\vec{v})
\left( 1 +{\cal O}(\lambda^{3/2})\right)\, .
\label{F1G1} 
\end{eqnarray}
As shown in the Appendix, in the combined limit the form factors $F_1$ and 
$G_1$ are equal up to corrections of order $\lambda$. Thus, only a single 
function describes the entire electroweak matrix element. We denote this 
function as $\Theta$ defined by,
\begin{equation}
\Theta \equiv F_1 = G_1 \left( 1 +{\cal O}(\lambda)\right) \, .
\label{Thetadef}
\end{equation}
This function should be distinguished from the Isgur-Wise universal function 
$\eta(w)$ \cite{HQ1,HQ2,HQ3,HQ4,HQ5}. The latter is a smooth function of the 
velocity transfer parameter $w$ in the pure heavy quark expansion. The 
function $\Theta$ is only defined for $|\delta \vec v| \sim \lambda^{3/4}$; 
for this range of the velocity transfer the function $\Theta$ can be 
determined as an expansion in powers of $\lambda^{1/2}$. As discussed below, 
$\Theta$ is not a smooth function of $|\delta \vec v|$ in the combined limit. 
However, another kinematic variable, $z$, can be introduced so that $\Theta$ 
is a smooth function of $z$ in the combined limit. 

Let us focus on the following matrix element:
\begin{equation}
\langle \Lambda_c (\vec{v}^\prime)|c^{\dag} b|\Lambda_b (\vec{v}) \rangle =
\Theta \, u_{c}^{\dag} (\vec{v}^\prime)u_b (\vec{v}) 
\left( 1 +{\cal O}(\lambda^{3/2})\right) \, .
\label{Theta}
\end{equation}
This matrix element can be determined using the collective wave functions of 
the initial and final heavy baryons and the effective operator corresponding 
to the operator $c^\dag b$ in the combined limit. The collective wave functions
are determined in the rest frame of the heavy baryon by the effective 
Hamiltonian in eq.~(\ref{HEFF}). At leading order, they are given in 
eq.~(\ref{WF}). The next-to-leading order corrections come from the
term $\alpha x^4 /4!$ treated perturbatively. A collective wave function of
the heavy baryon moving with velocity $\vec v$ can be determined using the 
boost operator. In the combined limit the effective boost operator acting on 
the heavy baryon state is \cite{hb2}:
\begin{equation}
B_{\vec v}=e^{-i (m_N+m_H) \vec X \cdot \vec v}
\left( 1 +{\cal O}(\lambda)\right) \, .
\label{B}
\end{equation}

As shown in the Appendix, the effective operator corresponding to $c^\dag b$
at leading and next-to-leading order in the combined expansion is
$h_{c}^{\dag (v)}h_{b}^{(v)}$, where the heavy quark field 
$h_{Q}^{(v)}$ is defined in eq.~(\ref{hH}). Moreover, in the combined limit
this effective operator can be expressed in terms of the collective operator
$\vec X_b$ defined in eq.~(\ref{PQXQ}) \cite{hb2}:
\begin{equation}
h_{c}^{\dag (v)}(\vec y) h_{b}^{(v)} (\vec y) =\delta^3 (\vec X_b-\vec y)
\left(1+{\cal O} (\lambda)\right) \, .
\label{JEFF}
\end{equation}

Using the effective operator (eq.~(\ref{JEFF})), the boost operator 
(eq.~(\ref{B})) and the collective wave functions of the ground state heavy 
baryons (eq.~(\ref{WF})), the form factor $\Theta$ at leading order
in the combined expansion is:
\begin{equation}
\Theta= {2 \sqrt{2}\, \mu_{b}^{3/8} \mu_{c}^{3/8} \over 
(\sqrt{\mu_b}+\sqrt{\mu_c})^{3/2}} 
\exp \left( -{m_{N}^{2}|\delta \vec v|^2 \over 2 
(\sqrt{\kappa\mu_b}+\sqrt{\kappa\mu_c})} \right ) 
\left(1 +{\cal O}(\lambda) \right ) \, ,
\label{FF}
\end{equation}     
where $|\delta \vec v| \sim \lambda^{3/4}$. This result agrees with one 
obtained using the bound state picture \cite{bst3,bst4}. 

\begin{figure}[ht]
\begin{center}
\epsfig{figure=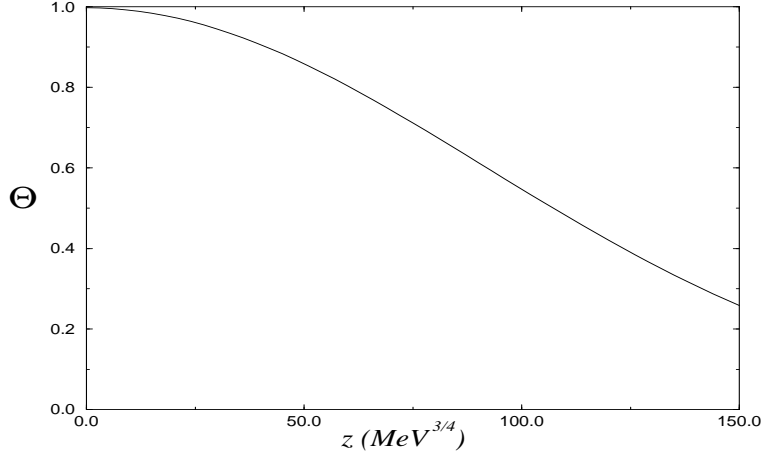,height=10cm,width=6cm,clip=,angle=-90}
\bigskip
\caption{The dominant semileptonic form factor for the
$\Lambda_b \to \Lambda_c \ell \bar{\nu}$ decay at LO defined in 
eq.~(\ref{Thetadef}) plotted as a function of the natural variable $z$
defined in eq.~(\ref{z}) with physical values of masses $m_N$, $\mu_c$ and 
$\mu_b$.}
\label{fig1}
\end{center}
\end{figure}

As a function of $|\delta \vec v|$, $\Theta$ has an essential singularity in 
the combined limit; it vanishes faster than any power of $\lambda$. However, 
as a function of the kinematic variable $z$ defined as,
\begin{equation}
z \equiv {m_{N}|\delta \vec v| \over 
(\sqrt{\mu_b}+\sqrt{\mu_c})^{1/2}} = {m_{N} \sqrt{2}(w-1)^{1/2} \over 
(\sqrt{\mu_b}+\sqrt{\mu_c})^{1/2}} \left (1+{\cal O}(\lambda^{3/2})\right)\,, 
\label{z}
\end{equation}
$\Theta$ is well behaved in the combined limit. For velocity transfers of 
order $\lambda^{3/4}$, variable $z$ can be  expressed in terms of a Lorentz 
scalar, $w$, since $w=(1+ |\delta \vec v|^2/2)\left(1 +{\cal O}
(\lambda^{3/2})\right)$. As a function of $z$, the form factor $\Theta$ at LO 
has the form:
\begin{equation}
\Theta \, (z) ={2 \sqrt{2}\,  
\mu_{b}^{3/8} \mu_{c}^{3/8} \over (\sqrt{\mu_b}+\sqrt{\mu_c})^{3/2}} 
\exp \left (-{ z^2 \over 2 \sqrt{\kappa}} \right )
\left(1 +{\cal O}(\lambda) \right ) \, .
\label{Thetaz}
\end{equation}     
The form factor $\Theta$ as a function of $z$ at LO is plotted in 
Fig.~\ref{fig1} for the physical values of the nucleon mass, $m_N$, and 
reduced masses, $\mu_c$ and $\mu_b$.

The experimentally measurable quantities are the value and derivatives of the 
form factors at zero recoil, or equivalently at $z=0$. The value of the form 
factor at zero recoil is given at LO by,
\begin{equation}
\Theta(z=0)={2 \sqrt{2} \, \mu_b^{3/8} \mu_c^{3/8} \over 
(\sqrt{\mu_b}+\sqrt{\mu_c})^{3/2}} \approx 0.998  \, .
\label{FF00}
\end{equation}
Thus, the form factor at zero recoil is completely determined at LO in the 
combined expansion in terms of the known masses, $\mu_c$ and $\mu_b$. 
In HQET, due to the heavy quark spin-flavor symmetry, the form factor at
zero recoil is unity up to corrections of order $1/m_{Q}^{2}$ \cite{mQ1}. The 
flavor symmetry is broken in the combined limit already at LO due to the 
dependence of kinetic energy of the collective motion on the flavor of the 
heavy quark via the reduced mass $\mu_Q$. However, the deviation from unity is
small for the $\Lambda_b \to \Lambda_c \ell \bar{\nu}$ decay. 

The second derivative of $\Theta (z)$ at zero recoil is given at LO in the 
combined limit by,
\begin{equation}
\rho \equiv {\partial^2 \Theta \over \partial^2 z}(z=0)  = 
{- 2 \sqrt{2}\, \mu_b^{3/8} \mu_c^{3/8} \over 
\sqrt{\kappa} (\sqrt{\mu_b}+\sqrt{\mu_c})^{3/2}} \left(1+{\cal O}(\lambda)
\right) \, .
\label{FF20}
\end{equation}
Expressing $\kappa$ in terms of the excitation energy of $\Lambda_c$, we get:
\begin{equation}
\rho= {- 2 \sqrt{2} \, \mu_b^{3/8} \mu_c^{-1/8} \over 
(m_{\bar{\Lambda}_{c}^{*}}- m_{\Lambda_{c1}})
(\sqrt{\mu_b}+\sqrt{\mu_c})^{3/2}} \approx -1.197\times 10^{-4}\,MeV^{-3/2}\,.
\label{ro}
\end{equation}

The NLO correction to the leading order expressions in eqs.~(\ref{FF}), 
(\ref{FF00}) and (\ref{ro}) are obtained by treating the next-to-leading 
order term in eq.~(\ref{HEFF}) perturbatively. The form factor $\Theta(z)$, 
its value and second derivative at zero recoil are given at NLO by,
\begin{eqnarray}
 & \Theta (z)= {2 \sqrt{2}\,  \mu_{b}^{3/8} \mu_{c}^{3/8} \over 
(\sqrt{\mu_b}+\sqrt{\mu_c})^{3/2}} \exp \left(-z^2 \over 2\sqrt{\kappa}\right) 
\nonumber \\
& \left (1+{\alpha \over 4! \kappa^{5/2} (\sqrt{\mu_b}+\sqrt{\mu_c})} 
\left ({45 \kappa(\sqrt{\mu_b}-\sqrt{\mu_c})^{2} \over 16 \kappa^{3/2} 
\sqrt{\mu_b\mu_c}} + {5 \, \kappa^{1/2} z^2} -{1\over 4} z^4 \right ) \right ) 
\left(1+{\cal O}(\lambda^{3/2})\right) \, , &
\label{F1NL0}
\end{eqnarray}

\begin{equation}
\Theta(z=0)  =  {2 \sqrt{2} \,\, \mu_{b}^{3/8} \mu_{c}^{3/8} \over 
(\sqrt{\mu_b}+\sqrt{\mu_c})^{3/2}}
\left ( 1+{\alpha \over 4!} {45 (\sqrt{\mu_b}-\sqrt{\mu_c})^{2} \over 
16 \kappa^{3/2}\sqrt{\mu_b\mu_c} (\sqrt{\mu_b}+\sqrt{\mu_c})} \right ) 
\left(1+{\cal O}(\lambda^{3/2})\right)  \, ,
\label{F10NLO}
\end{equation}

\begin{equation}
\rho = - {2 \sqrt{2} \,\, \mu_{b}^{3/8} \mu_{c}^{3/8} \over 
\sqrt{\kappa} (\sqrt{\mu_b}+\sqrt{\mu_c})^{3/2}}
\left ( 1+{\alpha \over 4!} {45 (\sqrt{\mu_b}-\sqrt{\mu_c})^{2} -
160 \sqrt{\mu_b \mu_c} \over 16 \kappa^{3/2}
\sqrt{\mu_b\mu_c} (\sqrt{\mu_b}+\sqrt{\mu_c})} \right ) 
\left(1+{\cal O}(\lambda^{3/2})\right)\, .
\label{F12NLO}
\end{equation}

As in the case with the spectroscopic observables in Sec.~\ref{S}, the 
semileptonic observables in eqs.~(\ref{F10NLO}) and (\ref{F12NLO}) are 
completely determined up to NNLO corrections in terms of two phenomenological
constants---$\kappa$ and $\alpha$.

\subsection{Decays of the Ground State to the Excited States of Heavy Baryons
\label{SDGE}}

There are two electroweak decay channels of the ground state of $\Lambda_b$
to the low-lying excited states of $\Lambda_c$: $\Lambda_b \to \Lambda_{c1}
\ell \bar{\nu}$ and $\Lambda_b \to \Lambda_{c1}^{*} \ell \bar{\nu}$. As will
be shown, at LO and NLO in the combined limit there is a single independent 
form factor that determines the hadronic matrix elements for each of the decay 
channels.
The approach here is similar to the one used in Sec.~\ref{SDGG}. The
$\lambda$-scaling of the form factors is determined by the self-consistent
counting rules of the hadronic matrix elements when the velocity transfer is 
of order $\lambda^{3/4}$.

The form factors for the $\Lambda_b \to \Lambda_{c1} \ell \bar{\nu}$ decay 
are given by, 
\begin{equation}
\langle \Lambda_{c1} (\vec{v}^\prime)|\bar c \gamma^\mu (1-\gamma_5) b |
\Lambda_b (\vec{v}) \rangle = 
\bar{u}_{c} (\vec{v}^\prime) \left (\Gamma_V - \Gamma_A \right ) 
u_b (\vec{v})\, ,
\label{ampl1} 
\end{equation}
where
\begin{eqnarray}
\Gamma_V & = & K_1 \gamma^\mu +i K_2 \sigma^{\mu\nu} q_\nu +
i K_3  \sigma^{\mu\nu} p_\nu \, ,
\nonumber \\ 
\Gamma_A & = & \left (L_1 \gamma^\mu +L_2 p^\mu +L_3 q^\mu \right)\gamma_5 \, ,
\label{GVGA1} 
\end{eqnarray}
so that the parameterization is the same as in eq.~(\ref{JNR}). 

The $\lambda$-scaling of the time and space components of the vector and 
axial current matrix elements in eq.~(\ref{ampl1}) is the same as for the 
corresponding amplitudes in $\Lambda_b \to \Lambda_{c}\ell \bar{\nu}$ decay 
eq.~(\ref{CRLHS}). Combining these scaling rules with the scaling of the 
factors containing Dirac spinors and Dirac matrices, eq.~(\ref{KCR}), we get:
\begin{eqnarray}
K_1 & \sim & \lambda^0 , \,\,\,\,\, K_2 \sim \lambda , \,\,\,\,\,
 K_3 \sim \lambda \, ,  
\nonumber \\
L_1 & \sim & \lambda^0 , \,\,\,\,\,  L_2 \sim  \lambda , \,\,\,\,\,
L_3  \sim \lambda \, .
\label{KLCR}
\end{eqnarray}
The vector and axial form factors $K_1$ and $L_1$ are equal up to
corrections of order $\lambda$ as are the form factors $F_1$ and $G_1$. Thus, 
the dominant decay matrix element is determined by the single form factor:
\begin{equation}
\langle \Lambda_{c1} (\vec{v}^\prime)|c^{\dag} b|\Lambda_b (\vec{v}) 
\rangle =
K_1 u_{c}^{\dag} (\vec{v}^\prime)u_b (\vec{v})
\left(1 +{\cal O}(\lambda^{3/2})\right) \, .
\label{K1def}
\end{equation}
This matrix element can be calculated at LO and NLO in the combined expansion.

The hadronic matrix element of the $\Lambda_b \to \Lambda_{c1}^{*} \ell 
\bar{\nu}$ decay is parameterized by eight form factors:
\begin{equation}
\langle \Lambda_{c1}^{*} (\vec{v}^\prime)|\bar c \gamma^\mu (1-\gamma_5) b |
\Lambda_b (\vec{v}) \rangle = 
\bar{u}_{c\nu} (\vec{v}^\prime) \left ( \Gamma_V - \Gamma_A \right ) 
u_b (\vec{v}) \, ,
\label{ampl2} 
\end{equation}
where 
\begin{eqnarray}
\Gamma_V & = & ( N_1 \sigma^{\nu\mu} + N_2 \gamma^\mu q^\nu  +
i N_3 \sigma^{\mu\rho} q_\rho q^\nu +i N_4 \sigma^{\mu\rho} p_\rho q^\nu )
\gamma_5 \, , 
\nonumber \\
\Gamma_A & = & M_1 g^{\mu\nu}  + M_2 \gamma^\mu q^\nu  +
M_3 q^\mu q^\nu  + M_4 q^\mu p^\nu  \, ,
\label{GVGA2}
\end{eqnarray}
and $u_{c}^{\nu} (\vec v, s)$ is the Rarita-Schwinger spinors corresponding to 
the heavy baryons with total spin $3/2$. They are normalized so that
$\bar{u}_{Q\nu} u_{Q}^{\nu}=-1$. The time component of $u_\nu$ is suppressed 
by $\lambda^{3/4}$ relative to the special components for the velocities
of order $\lambda^{3/4}$.

The $\lambda$-scaling for the velocity transfers of order $\lambda^{3/4}$
of the vector and axial matrix elements in eq.~(\ref{ampl2}) are the same 
as in eq.~(\ref{CRLHS}). The kinematic factors scale as:
\begin{eqnarray}
\bar{u}_{c\nu} (\vec{v}^\prime)\sigma^{\nu i} \gamma_5 u_b (\vec{v})
& \sim & \lambda^{3/4} 
\, ,  \nonumber \\
\bar{u}_{c\nu} (\vec{v}^\prime) \gamma^i q^\nu \gamma_5 u_b (\vec{v})
& \sim & \lambda^{-1/4} 
\, ,  \nonumber \\
\bar{u}_{c\nu} (\vec{v}^\prime) \sigma^{i\rho} q_\rho q^\nu \gamma_5 
u_b (\vec{v})
& \sim & \lambda^{-5/4} 
\, ,  \nonumber \\
\bar{u}_{c\nu} (\vec{v}^\prime) \sigma^{i\rho} p_\rho q^\nu \gamma_5 
u_b (\vec{v})
& \sim & \lambda^{-5/4}  
\label{GVVCR}
\end{eqnarray}
for the spatial components, and,
\begin{eqnarray}
\bar{u}_{c\nu} (\vec{v}^\prime) \sigma^{\nu 0} \gamma_5 u_b (\vec{v})
& \sim & \lambda^{0} 
\, ,\nonumber \\
\bar{u}_{c\nu} (\vec{v}^\prime) \gamma^0 q^\nu \gamma_5 u_b (\vec{v})
& \sim & \lambda^{1/2} 
\, ,  \nonumber \\
\bar{u}_{c\nu} (\vec{v}^\prime) \sigma^{0\rho} q_\rho q^\nu \gamma_5 
u_b (\vec{v})
& \sim & \lambda^{-1/2} 
\, ,  \nonumber \\
\bar{u}_{c\nu} (\vec{v}^\prime) \sigma^{0\rho} p_\rho q^\nu \gamma_5 
u_b (\vec{v})
& \sim & \lambda^{-1/2}  
\label{GV0CR}
\end{eqnarray}
for the time component.
Since the spatial components of the vector current matrix elements are
of order $\lambda^{3/4}$ (eq.~(\ref{CRLHS})), the scaling rules in 
eq.~(\ref{GVVCR}) lead to the following scaling rules of the
vector form factors:
\begin{eqnarray}
N_1 & \sim & \lambda^0 , \,\,\,\,\, N_2 \sim \lambda , \,\,\,\,\,
N_3 \sim \lambda^{2}, \,\,\,\,\, N_4 \sim \lambda^{2} \, .  
\label{NCR}
\end{eqnarray}
These scaling rules are consistent with the scaling of the time component
of the vector current amplitude (eqs.~(\ref{CRLHS}) and (\ref{GV0CR})).
Combining eqs.~(\ref{NCR}) and (\ref{GV0CR}) we get,
\begin{eqnarray}
N_1 \bar{u}_{c\nu} (\vec{v}^\prime) \sigma^{\nu 0} \gamma_5 u_b (\vec{v})
& \sim & \lambda^{0} 
\, ,  \nonumber \\
N_2 \bar{u}_{c\nu} (\vec{v}^\prime) \gamma^0 q^\nu \gamma_5 u_b (\vec{v})
& \sim & 
\lambda^{3/2}  \, ,  \nonumber \\
N_3 \bar{u}_{c\nu} (\vec{v}^\prime) \sigma^{0\rho} q_\rho q^\nu \gamma_5 
u_b (\vec{v})  
& \sim & \lambda^{3/2} 
\, ,  \nonumber \\
N_4 \bar{u}_{c\nu} (\vec{v}^\prime) \sigma^{0\rho} p_\rho q^\nu \gamma_5 
u_b (\vec{v})
& \sim & \lambda^{3/2} \, .
\label{NGV0CR}
\end{eqnarray}
Thus, only $N_1$ contributes to the vector current matrix element
at leading and next-to-leading order in the combined expansion. The
dominant matrix element is:
\begin{equation}
\langle \Lambda_{c1}^{*} (\vec{v}^\prime)|c^{\dag} b |
\Lambda_b (\vec{v}) \rangle = 
i N_1 \bar{u}_{c}^{j} (\vec{v}^\prime) \Sigma^j u_b (\vec{v}) +{\cal O}
(\lambda^{3/2}) =
i \sqrt{2 \over 3} N_1 \bar{u}_{c} (\vec{v}^\prime) u_b (\vec{v}) 
\left(1 +{\cal O}(\lambda^{3/2})\right)\, ,
\label{N1def} 
\end{equation}
where the expressions for the Rarita-Schwinger spinors in terms of 
Dirac spinors were used.

The scaling of the axial form factors is determined by the scaling of the 
time component of the axial current matrix element and by the 
scaling of the corresponding kinematic factors:
\begin{eqnarray}
\bar{u}_{c\nu} (\vec{v}^\prime) g^{0\nu} u_b (\vec{v})
& \sim & \lambda^{3/4} 
\, ,\nonumber \\
\bar{u}_{c\nu} (\vec{v}^\prime) \gamma^0 q^\nu u_b (\vec{v})
& \sim & \lambda^{-1/4} 
\, ,  \nonumber \\
\bar{u}_{c\nu} (\vec{v}^\prime) q^0 q^\nu u_b (\vec{v})
& \sim & \lambda^{-5/4} 
\, ,  \nonumber \\
\bar{u}_{c\nu} (\vec{v}^\prime) q^0 p^\nu u_b (\vec{v})
& \sim & \lambda^{-5/4} \, .
\label{GA0CR}
\end{eqnarray}
These scaling rules lead to the following scaling of the axial form factors:
\begin{eqnarray}
M_1 & \sim & \lambda^0 , \,\,\,\,\, M_2 \sim \lambda , \,\,\,\,\,
M_3 \sim \lambda^{2}, \,\,\,\,\, M_4 \sim \lambda^{2} \, .  
\label{MCR}
\end{eqnarray}
Hence, the terms on the right-hand side in eq.~(\ref{ampl2}) scale as,
\begin{eqnarray}
M_1 \bar{u}_{c\nu} (\vec{v}^\prime) g^{i\nu} \gamma_5 u_b (\vec{v})
& \sim & \lambda^{0} 
\, ,\nonumber \\
M_2 \bar{u}_{c\nu} (\vec{v}^\prime) \gamma^i q^\nu u_b (\vec{v})
& \sim & \lambda^{3/2} 
\, ,  \nonumber \\
M_3 \bar{u}_{c\nu} (\vec{v}^\prime) q^i q^\nu u_b (\vec{v})
& \sim & \lambda^{3/2} 
\, ,  \nonumber \\
M_4 \bar{u}_{c\nu} (\vec{v}^\prime) q^i p^\nu u_b (\vec{v})
& \sim & \lambda^{3/2} \, ,
\label{MGA0CR}
\end{eqnarray}
so that the dominant axial matrix element for $\Lambda_b \to \Lambda_{c1}^{*}
\ell \bar{\nu}$ decay is determined by a single form factor at LO and NLO
in the combined limit:
\begin{equation}
\langle \Lambda_{c1}^{*} (\vec{v}^\prime)|\bar{c} \gamma^i \gamma_5 b |
\Lambda_b (\vec{v}) \rangle = M_1 \bar{u}_{c}^i (\vec{v}^\prime) 
u_b (\vec{v}) +
{\cal O}(\lambda^{3/2}) =
\sqrt{2 \over 3} M_1 \bar{u}_{c}(\vec{v}^\prime) u_b (\vec{v}) \delta^{i3} +
{\cal O}(\lambda^{3/2}) \, .
\label{M1def} 
\end{equation}

As shown in the Appendix the form factors $N_1$ and $M_1$ are equal up to 
corrections of order $\lambda$. Hence, the hadronic matrix element of 
$\Lambda_b \to \Lambda_{c1}^{*}\ell \bar{\nu}$ decay is determined by a single 
independent form factor given by,
\begin{equation}
\langle \Lambda_{c1}^{*} (\vec{v}^\prime)|c^{\dag} b |
\Lambda_b (\vec{v}) \rangle = 
\sqrt{2 \over 3} N_1 u_{c}^{\dag} (\vec{v}^\prime) u_b (\vec{v}) 
\left(1 +{\cal O}(\lambda^{3/2})\right) \, .
\label{N1final} 
\end{equation}

The form factors $N_1$ and $K_1$ are not independent since states 
$|\Lambda_{c1} \rangle$ and $|\Lambda_{c1}^{*} \rangle$ are degenerate 
in the combined limit (they are related by the heavy quark spin flip) and the 
effective operator in eq.~(\ref{JEFF}) is independent of the heavy quark spin.
The boost operator and the effective operator in eq.~(\ref{JEFF}) connect 
states with definite orbital angular momentum $l$ and heavy quark spin $s_Q$. 
Using appropriate Clebsch-Gordan coefficients, the form factor $K_1$ is given
in terms of the form factor $N_1$ by,
\begin{equation}
K_1=\sqrt{1\over 3} N_1 \left(1 +{\cal O}(\lambda)\right) \, .
\label{K1N1}
\end{equation}

\begin{figure}[ht]
\begin{center}
\epsfig{figure=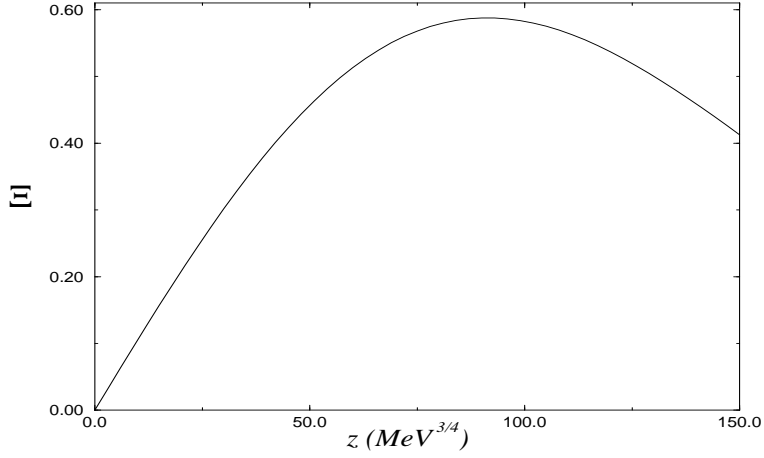,height=10cm,width=6cm,clip=,angle=-90}
\bigskip
\caption{The dominant semileptonic form factor for the 
$\Lambda_b \to \Lambda_{c1}^{*} (\Lambda_{c1}) \ell \bar{\nu}$
decays at LO defined in eq.~(\ref{Ksi}) plotted as a function of the 
natural variable $z$ defined in eq.~(\ref{z}) with physical values of masses 
$m_N$, $\mu_c$ and $\mu_b$.}
\label{fig2}
\end{center}
\end{figure}

As in the case of the $\Lambda_b \to \Lambda_c \ell \bar{\nu}$, the 
electroweak decays of the ground state of $\Lambda_b$ into the doublet
of the first orbitally excited state is determined by a single
form factor at LO and NLO in the combined limit:
\begin{equation}
\Xi \equiv N_1 = M_1 \left(1+{\cal O}(\lambda)\right) = 
\sqrt{3}\,  K_1  \left(1+{\cal O}(\lambda)\right) =
\sqrt{3} \, L_1  \left(1+{\cal O}(\lambda)\right) \, .
\label{Ksi}
\end{equation}
We denote this function as $\Xi$ to distinguish it from the universal heavy
quark form factor in HQET, Ref.~\cite{HQ1,HQ2,HQ3,HQ4,HQ5}, since it applies 
in the combined limit.
The function $\Xi$ can be calculated from eq.~(\ref{N1final}) using the 
effective boost operator in eq.~(\ref{B}), the heavy quark effective
operator in eq.~(\ref{JEFF}) and the collective wave functions of the ground 
and the first excited states of heavy baryons in eq.~(\ref{WF}). Hence, at 
leading order the function $\Xi$ is given by,
\begin{equation}
\Xi =  {4 \, m_N |\delta \vec v| \mu_{b}^{3/8} \mu_{c}^{5/8} \over 
\kappa^{1/4}(\sqrt{\mu_b}+\sqrt{\mu_c})^{5/2}}
\exp \left ({-{m_{N}^{2} |\delta \vec v|^2 \over 
2(\sqrt{\kappa \mu_b}+\sqrt{\kappa \mu_c})}} \right )
\left(1 +{\cal O}(\lambda^{3/2}) \right)\, ,
\label{KsiLO}
\end{equation} 
where the velocity transfer $|\delta \vec v|$ is of order $\lambda^{3/4}$.
The form factor can be expressed as a function of the kinematic variable
$z$ defined in eq.~(\ref{z}):
\begin{equation}
\Xi \, (z) = {4 \, z \, \mu_{b}^{3/8} \mu_{c}^{5/8} \over 
\kappa^{1/4}(\sqrt{\mu_b}+\sqrt{\mu_c})^{2}} \, 
\exp \left ({- z^2 \over 2 \sqrt{\kappa}} \right ) 
\left(1 +{\cal O}(\lambda^{3/2}) \right)\, .
\label{KsiLOz}
\end{equation} 
As a function of $z$, $\Xi (z)$ has a smooth expansion in powers of 
$\lambda^{1/2}$ near zero recoil. The function $\Xi (z)$ is plotted in 
Fig.~\ref{fig2}.

The slope of $\Xi$ at zero recoil at LO is given by,
\begin{eqnarray}
\sigma &\equiv& {\partial \Xi \over \partial z} \, (z=0) = 
{4 \,\,  \mu_{b}^{3/8} \mu_{c}^{5/8} \over 
\kappa^{1/4}(\sqrt{\mu_b}+\sqrt{\mu_c})^{2}} \left(1+{\cal O} (\lambda) 
\right)
\nonumber \\
&=& {4 \,\, \mu_{b}^{3/8} \mu_{c}^{3/8} \over 
\sqrt{m_{\bar{\Lambda}_{c}^{*}}- m_{\Lambda_{c1}}}
(\sqrt{\mu_b}+\sqrt{\mu_c})^{2}} \left(1+{\cal O} (\lambda) \right)
\approx 0.011 \, MeV^{-3/4}  \, , 
\label{Ksislope}
\end{eqnarray}
where in the third equality the constant $\kappa$ at LO is expressed in terms 
of the excitation energy of the first excited state of $\Lambda_c$.

At NLO the function $\Xi (z)$ and its slope at zero recoil are given by,  
\begin{eqnarray}
& \Xi \, (z) = {4 \, z \,  \mu_{b}^{3/8} \mu_{c}^{5/8} \over 
\kappa^{1/4}(\sqrt{\mu_b}+\sqrt{\mu_c})^{2}}  
\exp\left( -{z^2 \over 2 \sqrt{\kappa}} \right)&
\nonumber \\
& \left ( 1+{\alpha \over 4! \kappa^{5/2} (\sqrt{\mu_b}+\sqrt{\mu_c})}
\left ({(105 \mu_b - 230 \sqrt{\mu_b \mu_c}+ 45 \mu_c) \kappa \over 16 
\sqrt{\mu_b \mu_c}} + {13\over2}\, \kappa^{1/2} z^2 -{1\over 4}\, z^4\right ) 
\right ) \left(1+ {\cal O}(\lambda^{3/2}) \right )\, , &
\label{KsiNLO}
\end{eqnarray} 

\begin{equation}
\sigma = {4 \,\, \mu_{b}^{3/8} \mu_{c}^{5/8} \over 
\kappa^{1/4}(\sqrt{\mu_b}+\sqrt{\mu_c})^{2}}
\left ( 1+{\alpha \over 4!} 
{105 \mu_b - 230 \sqrt{\mu_b \mu_c}+ 45 \mu_c \over 16 \kappa^{3/2} 
\sqrt{\mu_b \mu_c} (\sqrt{\mu_b}+\sqrt{\mu_c})} \right )
\left (1+{\cal O}(\lambda^{3/2}) \right) \, .
\label{slopeNLO}
\end{equation}  
The slope is completely determined in terms of the constants $\kappa$ and
$\alpha$.

\section{Radiative decays of excited heavy baryons \label{RD}}

The effective theory near the combined heavy quark and large $N_c$ limit can 
be used to predict electromagnetic decay rates of excited heavy baryons. 
The strong decays of excited $\Lambda_{c}$ baryons are dominated by three-body
isospin conserving decays with two final pions. However, because of the small
energy splitting between these states and the ground state, the phase space 
for these decays is restricted. Due to this phase space restriction, 
the radiative transitions may even be the dominant decay mode in the bottom 
sector. Thus, the  strong decays are greatly 
suppressed and electromagnetic decays will have a substantial branching ratio.
The radiative decays of excited $\Lambda_c$ and $\Lambda_b$ baryons within the
framework of the bound state picture of heavy baryons were considered in 
\cite{HQ15}. Here we will use the effective theory to perform a model 
independent analysis up to NLO. The effective theory gives the form of the 
matrix elements of a dipole operator between the low-lying bound states of a 
heavy baryon which determine the leading order contribution to the decay 
amplitude. At leading order, the radiative decay rates are completely 
determined in terms of the constant $\kappa$. At next-to-leading order, an 
additional constant---$\alpha$---is needed. As will be shown, no additional 
phenomenological parameters associated with the electromagnetic current arise.
   
An electromagnetic decay amplitude is determined by the interaction 
Hamiltonian, ${\cal H}_{int}=e\,\int d^3 x j^{\mu}(\vec x) A_{\mu}(\vec x)$, 
where $j^{\mu}(\vec x)$ is a current operator that couples to a photon field
$A^{\mu}(\vec x)$ with coupling constant $e$. According to Fermi's golden 
rule, the decay rate is proportional to the square of the absolute value of
the matrix element of ${\cal H}_{int}$ between initial and final states. 
If the wavelength of a radiated photon is much larger than the typical 
size of the system, $\omega a \ll 1$ ($\omega$---the energy of a photon,
$a$---typical size), then the matrix element can be expanded in terms
of multipole operators \cite{HQ16}:
\begin{equation}
\langle f|{\cal H}_{int}|i \rangle = (-1)^{m+1} i^{J} \sqrt{{(2J +1)(J+1) 
\over \pi J}} 
{\omega^{J+ {1 \over 2}} \over (2J+1)!!} e ((M^{(e)}_{J, -m})_{fi}+
(M^{(m)}_{J, -m})_{fi}) , 
\label{HEM}
\end{equation} 
where the initial state consists of a single heavy baryon in an excited state;
the final state consists of a heavy baryon in the ground state and a photon 
with a definite 
angular momentum, $J$, and its $z$-component, $m$. The matrix elements 
$(M^{(e)}_{J, m})_{fi}$ and $(M^{(m)}_{J, m})_{fi}$ correspond to electric,
$E_J$, and magnetic, $M_J$, transitions, respectively; they are given by,
\begin{equation}
\left (M^{(e)}_{J, m} \right )_{fi}=\sqrt{{4 \pi \over (2J+1)}} \int d^{3}x 
\rho_{fi}(\vec x)
|\vec x|^{J} Y_{Jm}({\vec x \over |\vec x|}),
\label{mmE}
\end{equation} 
\begin{equation}
\left (M^{(m)}_{J, m} \right )_{fi}={1 \over (J+1)} 
\sqrt{{4 \pi \over (2J+1)}} \int d^{3}x 
(\vec x \times {\vec j}_{fi}(\vec x)) \cdot 
\nabla (|\vec x|^{J} Y_{Jm}({\vec x \over |\vec x|})),
\label{mmM}
\end{equation}
where $(\rho_{fi}(\vec x), {\vec j}_{fi}(\vec x))= j^{\mu}_{fi}(\vec x)$.

Fermi's golden rule gives the total decay rates in which photons with given 
values of $J$ and $m$ are radiated:
\begin{equation}
\Gamma^{(e)}={2 (2J+1)(J+1) \over J ((2J+1)!!)^2} {\omega^{2J+1}} e^2
{\mid (M^{(e)}_{J,-m})_{fi} \mid}^2
\label{GE}
\end{equation} 
for $E_J$ transitions, and,
\begin{equation}
\Gamma^{(m)}={2 (2J+1)(J+1) \over J ((2J+1)!!)^2} {\omega^{2J+1}} e^2
{\mid (M^{(m)}_{J,-m})_{fi} \mid}^2
\label{GM}
\end{equation}
for $M_J$ transitions.  

In order to arrive at eqs.~(\ref{mmE}) and (\ref{mmM}), only the first term
in the expansion of the photon radial wave functions is kept. These wave 
functions are spherical Bessel functions, $g_{J}(k|\vec x|), which equal to
(\pi /2k|\vec x|)^{1/2} J_{J+1/2}(k|\vec x|)$, where $J_{J+1/2}(k|\vec x|)$ is
the Bessel function of the first kind. This is valid when $k|\vec x| 
\sim ka \ll 1$. The next-to-leading order term in the expansion of the radial 
wave function is suppressed by $(k|\vec x|)^2$. For the decay of the 
first excited state of a heavy baryon to its ground state, this condition is 
satisfied in the combined limit. From the counting rules of our effective
theory the momentum $k=\omega \sim \lambda^{1/2}$, and the typical size is 
determined by the expectation value of the operator $\vec x$, which is of  
order $\lambda^{1/4}$. Thus, $(k|\vec x|)^2 \sim \lambda^{3/2} \ll 1$ near the
combined limit validating the multipole expansion. In this approximation, 
decay rates corresponding to higher 
order $E_J$ and $M_J$  operators are suppressed by $(k|\vec x|)^2 \sim 
\lambda^{3/2}$ (as is evident from eqs.~(\ref{mmE}) and (\ref{mmM})). The 
$M_J$ decay rate is suppressed relative to the $E_J$ decay with 
the same $J$ by ${\mid \vec v \mid}^2 \sim \lambda^{3/2}$. In addition, the 
transitions between states with definite total angular momentum can only occur
either through $E_J$ or $M_J$ decays due to parity conservation in the 
electromagnetic interaction; $E_J$ photons have parity $(-1)^J$, while 
$M_J$ photons have parity $(-1)^{J+1}$. Hence, at LO and NLO in the combined
limit the dominant mode for the elecromagnetic decay of the first excited 
state of a heavy baryon (with negative parity) into the ground state (with 
positive parity) is the $E1$ decay. It is determined by the matrix element of 
the dipole operator of the heavy baryon, $\vec d$. Summing eq.~(\ref{GE}) 
over all possible values of $m$ for $J=1$, $\it i.e.\, m=0, \pm 1$, we get 
the decay rate,
\begin{equation}
\Gamma= {4 \omega^3 \over 3} {|{\vec d}_{fi}|}^2,
\label{TDR}
\end{equation}
with the dipole matrix element defined as,
\begin{equation}
{\vec d}_{fi}=\, e \,\int d^3 x \rho_{fi}(\vec x) \vec x.
\label{dfi}
\end{equation}
When recoil is taken into consideration, eq.~(\ref{TDR}) is multiplied by 
$(1-\omega/2M)$, where $M$ is the total mass of the initial baryon. However, 
because this correction is of relative order $\lambda^{3/2}$, it can 
be neglected.
  
The combined expansion can be used to determine the dipole matrix 
element in eq.~(\ref{dfi}). The charge density in eq.~(\ref{dfi}) is the time
component of the current $j^{\mu}$: 
\begin{equation}
{\vec d}_{fi}= \langle \,f \, | e_Q \int d^3 x \,  Q^{\dag}(x) \, \vec x \,
Q (x) \, |\,  i \, \rangle +
\langle \, f\, | \, \sum_{i=u,d} e_i \int d^3 x \, q_{i}^{\dag}(x)\, 
\vec x \, q_{i} (x) \, |\, i \,\rangle \, ,
\label{d}
\end{equation}
where $e_Q$ and $e_i$ are the heavy quark and light quark charges in units of 
$e$; the sum is over the light quark flavors $u$ and $d$. The first term in 
eq.~(\ref{d}) is the matrix element of heavy quark position operator in the 
combined limit, eq.~(\ref{PQXQ}), times the heavy quark charge. The second 
matrix element---the brown muck contribution to the total dipole moment---can 
be written as,
$${\vec d}_{\ell} \equiv \langle \, f\, | \, {\sum_{i=u,d}}
(I_{3i}+{B_i \over 2}) \int d^3 x \, q_{i}^{\dag}(x)\, \vec x \, 
q_{i} (x) \, |\, i \,\rangle  $$
\begin{equation}
= \langle \, f\, | \, \sum_{i=u,d} \, I_{3i} \,
\int d^3 x \, q_{i}^{\dag}(x)\, \vec x \, 
q_{i} (x) \, |\, i \,\rangle 
+\langle \, f\, | \, \sum_{i=u,d} \, {B_i \over 2} \,
\int d^3 x \, q_{i}^{\dag}(x)\, \vec x \, 
q_{i} (x) \, |\, i \,\rangle \, , 
\label{dlight}
\end{equation}
where $I_{3i}$ is the third component of the isospin and $B_i$ is the baryon
number of a quark which is the same for every flavor. The first term in 
eq.~(\ref{dlight}) vanishes since it is an isospin one operator between 
isospin zero states. The baryon number of a single quark can be taken to be 
$1/N_c$, so that the total baryon number of the heavy baryon is still unity in
the combined limit. Thus, the light quark contribution to the dipole moment is:
\begin{equation}
{\vec d}_{\ell}={1 \over 2 N_c} \langle \, f\, | \,\sum_{i=u,d}  \,
\int d^3 x \, q_{i}^{\dag}(x)\, \vec x \, 
q_{i} (x) \, |\, i \,\rangle \, .
\label{dl}
\end{equation} 

In analogy to the heavy quark part, one would expect that the brown muck 
contribution to the the dipole moment is proportional to the operator 
$\vec X_\ell$. However, the operator in eq.~(\ref{dl}) contains only quark
fields while $\vec X_\ell$ acts on gluon fields as well. Nevertheless, as was 
shown in Ref.~\cite{HQ1}, in the subspace of the low-lying states the dipole 
moment of the brown muck is proportional to ${\vec X}_\ell$ with 
proportionality constant completely determined up to corrections of order 
$\lambda$:
\begin{equation}
{1 \over N_c -1} \langle f | \sum_{i=u,d}
\, \int d^3 x \, q^{\dagger}_{i}\,{\vec x} \, q_{i} | i \rangle  =
\langle f |{\vec X}_\ell | i \rangle \left(1+{\cal O}(\lambda)\right).
\label{effdl}
\end{equation}
Combining eqs.~(\ref{d}), (\ref{dlight}), (\ref{dl}) and (\ref{effdl}), we get:
\begin{equation}
{\vec d}_{fi}=\, e \, \langle f | e_{Q}\, {\vec X}_{Q}+ e_{\ell}
{\vec X}_{\ell} | i \rangle \left ( 1+{\cal O}(\lambda) \right )\, ,
\label{dfi3}
\end{equation}
where the effective brown muck charge is given by,
\begin{equation}
e_{\ell}=(N_c-1)/2 N_c \, .
\label{el}
\end{equation}
To get the correct total charges for $\Lambda_c$ and $\Lambda_b$ baryons,
the heavy quark charge is given by,
\begin{equation}
e_Q=(1 \pm N_c)/2N_c 
\label{eQ}
\end{equation} 
for $\Lambda_c$ and $\Lambda_b$ baryons, respectively.

The charge assignment for the heavy quark and the brown muck is somewhat 
ambiguous. One possibility is to use the same charge assignment as for 
$N_c = 3$, namely $e_{Q=c}=+2/3$, $e_{Q=b}=-1/3$, and for the brown muck 
$e_{\ell}=+1/3$. Another way is to choose an assignment valid in the combined
limit: 
\begin{equation} 
e_Q= \pm 1/2 \left(1+{\cal O}(\lambda)\right), \,\,\,\, 
e_\ell =1/2 \left(1+{\cal O}(\lambda)\right) \, .
\label{LNcharge}
\end{equation}
This charge assignment is used in the bound state picture of the heavy baryon,
where the heavy meson and the nucleon are in the superposition of two
isospin states \cite{HQ15}. The ambiguity between two charge assignments
is of order $\lambda$ and hence can not be resolved  at the order to which
we are working. The charge assignment in eq.~(\ref{LNcharge}) is also obtained
by considering anomaly cancelation of the $SU(N_c) \times SU(2) \times U(1)$ 
generalization of the standard model \cite{charge1,charge2}.

Using eqs.~(\ref{X}) and (\ref{xp}), the dipole matrix element in 
eq.~(\ref{dfi3}) can be expressed in terms of the relative position operator 
$\vec x$:
\begin{equation}
{\vec d}_{fi}=\, e \, {\mu_Q} ({e_l \over m_N}-{e_Q \over m_Q}) 
\langle f | {\vec x} | i \rangle \, .
\label{dfi4}
\end{equation}
Note, this operator is identical to the one used in the bound state picture
\cite{HQ15}. However, in the present context it emerges from QCD in 
the combined limit without additional assumptions. 

The dipole matrix element between the first excited state and the ground state
in eq.~(\ref{dfi4}) can be evaluated up to NLO using the collective wave
functions discussed in Sec.~\ref{S}. The decay rate is given by 
eq.~(\ref{TDR}). To find a total decay rate one needs to sum over all final 
states and average over the initial states. This gives the same result for
both $\Lambda_{Q1} \to \Lambda_{Q} \gamma$ and 
$\Lambda_{Q1}^{*} \to \Lambda_{Q} \gamma$ decays. At order $\lambda^{1/2}$ 
the total decay rate is completely determined in terms of the constant 
$\kappa$:
\begin{equation}
\Gamma(\Lambda_{Q1} \to \Lambda_Q \, \gamma)=\, {2\over 3} e^2 \kappa
\left ( {e_l \over m_N}-{e_Q \over m_H} \right )^2 
\left (1+ {\cal O}(\lambda) \right ) \, .
\label{TDR2}
\end{equation}
As was shown in Sec.~\ref{S}, at LO $\kappa$ is equal to the square of the
excitation energy of $\Lambda_{c1}$ times the reduced mass, 
$\kappa= \omega_{Q}^{2} \mu_Q=
(m_{\bar{\Lambda}_{c}^{*}}-m_{\Lambda_c})^2 \mu_c$. Using the charge 
assignment in eq.~(\ref{LNcharge}), the total decay rates are given by:
\begin{equation}
\Gamma(\Lambda_{c1} \to \Lambda_c \, \gamma)=\, {1\over 6} e^2 
{(m_{\bar{\Lambda}_{c}^{*}}-m_{\Lambda_c})^2 \over m_{\bar{D}} m_{N}}
{(m_{\bar{D}}-m_N)^2 \over m_{\bar{D}}+m_N } \approx 0.025 \, MeV \,
\left (1+ {\cal O}(\lambda) \right ) \,, 
\label{DRc}
\end{equation}
\begin{equation}
\Gamma(\Lambda_{b1} \to \Lambda_b \, \gamma)=\, {1\over 6} e^2 
{m_{\bar{D}} (m_{\bar{\Lambda}_{c}^{*}}-m_{\Lambda_c})^2   
\over  m_{\bar{B}}^{2} m_{N}}
{(m_{\bar{B}}+m_N)^2  \over m_{\bar{D}} + m_N} 
\approx 0.130 \, MeV \, \left (1+ {\cal O}(\lambda) \right )\,.
\label{DRb}
\end{equation}

It is interesting to consider the ratio of the decays in eqs.~(\ref{DRc}) and
(\ref{DRb}): 
\begin{equation}
{\Gamma(\Lambda_{c1}\to \Lambda_c \,\gamma) \over \Gamma(\Lambda_{b1}\to 
\Lambda_b \, \gamma)}={(m_D-m_N)^2 \over (m_B+m_N)^2}\, {m_{B}^2 
\over m_{D}^2} \approx 0.2 \, \left (1+ {\cal O}(\lambda) \right )\, .
\label{ratio}
\end{equation}
This is striking since in the pure heavy quark limit this ratio is unity. In 
this limit, the heavy quark is located at the center of mass of the heavy 
baryon. As a result, the dipole moment is determined by the motion of the 
brown muck alone relative to the center-of-mass. The pure heavy quark 
expansion is valid  if the recoil of the heavy quark due to its finite mass is 
suppressed (it is an order $1/m_Q$ effect). Thus, one might expect the ratio 
in eq.~(\ref{ratio}) to be close to $1$. However, if the mass of the brown 
muck is not much smaller than the mass of the heavy quark, as is the case for 
$\Lambda_c$ baryons and their excited states, the recoil of the heavy 
quark is not greatly suppressed. As a result, the motion
of the heavy quark relative to the center of mass of a heavy baryon gives a
significant contribution to the dipole moment of the system. The contributions
to the dipole moment due to the heavy quark and the brown muck motions 
around the center of mass partially cancel when their respective charges have 
the same sign. If the charges have opposite signs, the  two contributions add
together.
The partial cancelation happens for $\Lambda_c$ baryons, while for $\Lambda_b$ 
baryons the two contributions add \cite{HQ15}. This increases the suppression 
of the ratio of the decay rates in eq.~(\ref{ratio}). 

It is a general feature of effective theories that a nonperturbative
expansion (such as the $1/m_Q$ expansion in HQET) can work well for a number 
of observables, although it may have a very slow convergence for others. 
The failure of the pure heavy quark expansion of radiative decay rates of 
heavy baryons can be traced to a particular type of correction, {\it i.e.}
$m_N/m_H$, which are suppressed in the pure heavy quark limit but are of order
$\lambda^{0}$ in the combined limit. Phenomenologically, the ratio $m_N/m_H$ 
is not very small for heavy baryons, particularly for charm baryons. One would
like to sum these types of corrections to all orders to improve the 
convergence. This is accomplished by using the combined expansion. 

At next-to-leading order in the combined expansion, an additional 
constant---$\alpha$---is needed to determine the total decay rate. Once this 
constant is fixed from the spectroscopic or the semileptonic observables, the 
total decay rate can be determined by treating perturbatively the term of 
order $\lambda$ in the effective Hamiltonian (eq.~(\ref{HEFF})). The total 
decay rate at NLO is given by,
\begin{equation}
\Gamma(\Lambda_{c1} \to \Lambda_c \, \gamma) =
{1 \over 6 } e^2 \kappa {\left (m_{\bar{D}}-m_N \over m_{\bar{D}}m_N \right)^2}
\left  (1- {\alpha \over 4!} {5 \over \sqrt{\kappa^3 \mu_c}} \right )
\left (1+{\cal O}(\lambda) \right) \, ,
\label{results2}
\end{equation}
where the correction of order $\lambda$ arises from the ${\cal O}(\lambda)$
ambiguity in the charge assignment. A similar expression is obtained for the 
$\Lambda_{b1}$ decay:
\begin{equation}
\Gamma(\Lambda_{b1} \to \Lambda_b \, \gamma) =
{1\over 6 }e^2 \kappa {\left (m_{\bar{B}}+m_N \over m_{\bar{B}}m_N \right )^2}
\left (1- {\alpha \over 4!} {5 \over \sqrt{\kappa^3 \mu_b}}\right )
\left (1+{\cal O}(\lambda) \right) \, . 
\label{results3}
\end{equation}
Conversely, by measuring the radiative decay rates of the first excited states
of $\Lambda_c$ and $\Lambda_b$ baryons, the constants $\kappa$ and $\alpha$ 
can be fixed and used to predict the spectroscopic and semileptonic 
observables.

\section{Conclusion \label{C}}

We have studied the phenomenology of isoscalar heavy baryons within the
framework of an effective theory based on the contracted $O(8)$ in the 
combined heavy quark and large $N_c$ limit. This symmetry 
emerges in the subspace of the QCD Hilbert space with baryon number one and
heavy quark number one, {\it i.e} the heavy baryon subspace. The 
low-energy excited states of heavy baryons described by the effective theory 
are the collective excitations of the brown muck relative to the heavy quark. 

The effective Hamiltonian, eq.~(\ref{HEFF}), describes the low-lying 
excited states of isoscalar heavy baryons with excitation energies of order 
$\lambda^{1/2}$. Reliable predictions can be expected only for the ground 
state and a doublet of the first orbitally excited state. At leading 
nontrivial order, the spin-averaged sum of the first excited state of 
$\Lambda_b$ is predicted to be approximately $5920 \, MeV$. The 
next-to-leading corrections to the excitation energies, eq.~(\ref{P4}), 
are found by treating perturbatively the $\alpha x^4 /4!$ term in the 
effective Hamiltonian eq.~(\ref{HEFF}).
  
In addition to the spectroscopic observables, dominant semileptonic form
factors of the electroweak decays of $\Lambda_c$ are determined at LO and NLO.
We have shown, that at leading and next-to-leading order in the combined 
expansion there are only two independent form factors: one for 
$\Lambda_b \to \Lambda_c \ell \bar{\nu}$ transition, and one for 
$\Lambda_b \to \Lambda_{c1} \ell \bar{\nu}$ and 
$\Lambda_b \to \Lambda_{c1}^{*} \ell \bar{\nu}$ decays.
These form factors are calculated for the velocity transfers
of order $\lambda^{3/4}$. The form factors are exponentially suppressed for
velocity transfers of order unity. 

Experimentally useful quantities include values and derivatives of the form 
factors at zero recoil. We have determined these observables at leading 
and next-to-leading order. At LO, the form factor for $\Lambda_b \to 
\Lambda_c \ell \bar{\nu}$ at zero recoil is $0.998$ which is very close to 
the HQET normalization of unity. In the combined limit, the heavy quark is 
no longer the static source of the color magnetic field (in the baryon rest 
frame). As a result, we expect a deviation from the HQET result for the form 
factor normalization. However, as seen here this effect is very small.

We have calculated the total radiative decay rates of the excited heavy 
baryons. As we have shown, at LO and NLO in the combined expansion, these 
decays are dominated by the dipole radiation. There is a significant 
deviation of the ratio of these decay rates ($0.2$ at LO in the combined 
limit) from the HQET value of unity (eq.~(\ref{ratio})). 

At leading order (${\cal O}(\lambda^{1/2})$) our predictions are the same as 
those obtained in the bound state picture of heavy baryons 
\cite{bst1,bst2,bst3,bst4,HQ15}. However, our predictions are model 
independent and based on an effective theory with self-consistent counting 
rules. In addition, we have extended the treatment to next-to-leading order 
corrections which appear at relative order $\lambda^{1/2}$ and not at order 
$\lambda$ as was previously thought. In subsequent publications, we will 
extend the treatment of the effective theory to the excited states of non-zero
isopsin heavy baryons, such as $\Sigma_{c}^{*}$ and $\Sigma_{b}^{*}$. This 
can be done by combining the contracted $O(8)$ symmetry with the $SU(2N_f)$ 
symmetry which emerges in QCD in the large $N_c$ limit.

\acknowledgements
This work is supported by the U.S.~Department of Energy grant 
DE-FG02-93ER-40762.

\appendix
\section*{}

In this appendix we show that only one effective heavy quark operator 
contributes to the electroweak matrix element (eqs.~(\ref{JNR}), 
(\ref{ampl1}) and (\ref{ampl2})) at LO and NLO in the combined expansion. 
This operator is the first term in the combined expansion of the left-handed
heavy quark current. We also obtain the $\lambda$-scaling of the Lorentz
components of the effective operator which determine the self-consistent 
$\lambda$-scaling of the electroweak form factors discussed in Sec.~\ref{SD}.

The combined expansion of the heavy quark current 
$J=\bar{Q}_j (y)\Gamma Q_i (y)$ was developed in Ref.~\cite{hb2}. 
The expansion is achieved by separating the total heavy quark field $Q(y)$ 
into two parts:
\begin{eqnarray}
h_{Q}^{(v)}(y) &  = &  e^{-i(m_Q+m_N)\, v^\mu y_\mu }\, P_{+} \, Q(y) \, , 
\nonumber \\ 
H_{Q}^{(v)}(y) &  = & e^{-i(m_Q+m_N)\, v^\mu y_\mu}\, P_{-} \, Q(y) \, ,
\label{hH}
\end{eqnarray}
where the projection operators $P_{+}$ and $P_{-}$ are defined as 
$P_{+}=(1+\!\not\!\!v)/2$ and $P_{-}=(1-\!\not\!\!v)/2$. The fields 
$h_{Q}^{(v)}$ and $H_{Q}^{(v)}$ satisfy the following conditions:
\begin{equation}
\!\not\!\!v h_{Q}^{(v)}=h_{Q}^{(v)}, \,\,\,\, 
\!\not\!\!v H_{Q}^{(v)}=-H_{Q}^{(v)} \, .
\label{conditions}
\end{equation}

In HQET, the contribution of the effective field $H_{Q}^{(v)}$ is suppressed
by $1/m_Q$ relative to the field $h_{Q}^{(v)}$. The process of the heavy quark
pair creation is suppressed at leading order in HQET. We certainly expect the 
same in the combined limit. While, the heavy quark is bound to the massive 
brown muck, the heavy quark is still nonrelativistic at LO and NLO. Hence, 
the field $H_{Q}^{(v)}$ should be suppressed in the combined limit as well. 
However, it is not immediately clear due to the phase redefinition in 
eq.~(\ref{hH}). The heavy quark part of the total Lagrangian density can be 
expressed in terms of $h_{Q}^{(v)}$ and $H_{Q}^{(v)}$ using the conditions in 
eq.~(\ref{conditions}):
\begin{eqnarray}
{\cal L}_Q & = & \bar{Q}(i\!\not\!\! D -m_Q)Q  
 =  \bar{h}_{Q}^{(v)} \,( iv^\mu D_\mu)\,h_{Q}^{(v)} - 
\bar{H}_{Q}^{(v)} \,(iv^\mu D_\mu +2m_Q)\, H_{Q}^{(v)} + 
\bar{h}_{Q}^{(v)}\,( i{\!\not\!\! D}_\perp) \,H_{Q}^{(v)} 
\nonumber \\
& + & \bar{H}_{Q}^{(v)} \,(i{\!\not\!\! D}_\perp)\, h_{Q}^{(v)} + 
m_N \left (\bar{h}_{Q}^{(v)}\,h_{Q}^{(v)} - \bar{H}_{Q}^{(v)}\,H_{Q}^{(v)} 
\right ) \, , 
\label{LQCOMB}
\end{eqnarray}
where $D_{\perp}=D- (v^\nu D_\nu)v$ is the ``transverse part'' of the 
covariant derivative $D$.
The heavy quark Lagrangian in eq.~(\ref{LQCOMB}) written in terms of the 
fields $h_{Q}^{(v)}$ and $H_{Q}^{(v)}$ differs from its analog in HQET due 
to the additional term proportional to $m_N$.  This $m_N$ term would indicate
that the fields $h_{Q}^{(v)}$ and $H_{Q}^{(v)}$ are both heavy 
with masses $m_N$ and $m_Q+m_N$, respectively, which apparently prevents 
integrating out the ``heavy'' field $H_{Q}^{(v)}$. 

To get the correct scaling of the fields $h_{Q}^{(v)}$ and $H_{Q}^{(v)}$
one needs to consider the total QCD Lagrangian density which includes
contributions from the light degrees of freedom:
\begin{equation}
{\cal L}={\cal L}_Q+{\cal L}_q+{\cal L}_{YM}=\bar{Q}(i\!\not\!\! D - m_Q)Q+ 
\sum_j \bar{q}_{j}(i\!\not\!\! D -m_j)q_j +{\cal L}_{YM} \, ,
\label{LT}
\end{equation}
where the sum is over all light quarks, and ${\cal L}_{YM}$ is the Yang-Mills 
Lagrangian density. The total Lagrangian density in eq.~(\ref{LT}) can be 
re-expressed so as to build the brown muck contribution into the heavy degrees 
of freedom. In the Hamiltonian, eq.~(\ref{HEFF}), it is done by adding and 
subtracting a quantity which is an overall constant $m_N$ and regrouping terms
according to their $\lambda$ counting scaling. In the Lagrangian formalism 
this can be accomplished by using a Lorentz covariant operator 
$m_N \bar{Q}\!\not\!v Q$:
\begin{equation}
{\cal L}=({\cal L}_Q- m_N\bar{Q} \!\not\! v Q)+
({\cal L}_{q}+m_N \bar{Q}\!\not\!v Q 
+{\cal L}_{YM})={\cal L}_H+{\cal L}_\ell \, ,
\label{LTCOMB}
\end{equation}
where
\begin{eqnarray}
{\cal L}_H & \equiv & {\cal L}_Q- m_N\bar{Q}\!\not\! v Q \, ,
\nonumber \\
{\cal L}_\ell &\equiv & {\cal L}_{q}+m_N\bar{Q}\!\not\! v Q +{\cal L}_{YM} \,.
\label{LHLl}
\end{eqnarray}
In the rest frame of a heavy baryon operator, $\bar{Q} \!\not\!v Q$ is equal 
to the heavy quark density operator $Q^\dag Q$ and its spatial integral equals
to $1$ in the Hilbert space of heavy baryons. The heavy quark dynamics in the
combined limit is determined by ${\cal L}_H$. The additional mass term (with
minus sign) corresponds to the brown muck mass, the $m_N$ term in the effective
Hamiltonian, eq.~(\ref{HEFF}).

In terms of the fields $h_Q^{(v)}$ and $H_Q^{(v)}$, the operator 
$m_N \bar{Q}\!\not\!v Q$ is given by,
\begin{equation}
m_N\bar{Q}\!\not\!vQ=m_N \left ( \bar{h}_{Q}^{(v)}\,h_{Q}^{(v)}-
\bar{H}_{Q}^{(v)}\,H_{Q}^{(v)} \right ) \, .
\label{mNQQ}
\end{equation}
This operator cancels the last term in eq.~(\ref{LQCOMB}), so that the 
Lagrangian density ${\cal L}_H$ has the form:
\begin{equation}
{\cal L}_H=
\bar{h}_{Q}^{(v)} \,( iv^\mu D^\nu)\,h_{Q}^{(v)} - 
\bar{H}_{Q}^{(v)} \,(iv^\mu D^\nu +2m_Q)\, H_{Q}^{(v)} + 
\bar{h}_{Q}^{(v)}\,( i{\!\not\!\! D}_\perp ) \,H_{Q}^{(v)} +
\bar{H}_{Q}^{(v)} \,(i{\!\not\!\! D}_\perp)\, h_{Q}^{(v)}\, .
\label{LH}
\end{equation}
Hence, the ``large'' component $h_{Q}^{(v)}$ describes a massless field while 
the ``small'' component $H_{Q}^{(v)}$ has the mass $2m_Q$. Using equations of 
motion the ``heavy'' field $H_{Q}^{(v)}$ can be expressed in terms of the 
``light'' field $h_{Q}^{(v)}$ as follows:
\begin{equation}
H_{Q}^{(v)}= {1 \over 2m_Q+iv^\mu D_\mu}i \,{\!\not\!\! D}_\perp h_{Q}^{(v)}
\, .
\label{HQ}
\end{equation}
The relation between fields $h_{Q}^{(v)}$ and $H_{Q}^{(v)}$ 
in eq.~(\ref{HQ}) is identical in form to that in HQET between the ``large''
and ``small'' components of the heavy quark field $Q$. 

We can now expand the field 
$Q=e^{i(m_Q+m_N)v^\mu y_\mu} (h_{Q}^{(v)}+H_{Q}^{(v)})$, the Lagrangian 
density in eq.~(\ref{LH}), and the current $J=\bar{Q}_j \Gamma Q_i$, in powers 
of $\lambda$. This can be done by eliminating fields $H_{Q}^{(v)}$ (using 
eq.~(\ref{HQ})) and expanding the numerator in powers of $(iv^\mu D_\mu)/2m_Q$.
The combined expansion of the heavy quark field including terms up to 
order $\lambda$ is,
\begin{equation}
Q(y)=e^{-i(m_Q+m_N)v^\mu y_\mu} 
\left [ 1+ \left ({1-\!\not\!v \over 2} \right )
{i\!\not\!\!D \over 2m_Q} \right ] h_{Q}^{(v)}(y)\, + {\cal O}(\lambda^2).
\label{QCOMB}
\end{equation}
Using this expansion, the Lagrangian density ${\cal L}_H$ including terms up 
to order $\lambda$ has the form:
\begin{equation}
{\cal L}_H = \bar{h}_{Q}^{(v)} \,(i v^\mu D_\mu) \,h_{Q}^{(v)} 
 + {1 \over 2m_Q} \bar{h}_{Q}^{(v)} \left [-(i v^\mu D_\mu )^2+(iD)^2 -
{1\over2}g_s \sigma_{\mu \nu}G^{\mu \nu} \right ]\, h_{Q}^{(v)} + 
{\cal O}(\lambda^2),
\label{LHCOMB}
\end{equation}  
where $g_s$ is the strong coupling constant and 
$G^{\mu\nu}=[D^\mu , D^\nu ]$ is the gluon field strength tensor.
In a similar way, we can expand the current $J=\bar{Q}_j \Gamma Q_i$;  
keeping terms up to $\lambda$, we get,
\begin{equation}
J(y=0)= \bar{h}_{Q_j}^{(v)}\Gamma h_{Q_i}^{(v)} + 
{1 \over 2m_{Q_i}} \bar{h}_{Q_j}^{(v)}\Gamma (i\!\not\!\! D)
h_{Q_i}^{(v)} + {1 \over 2m_{Q_j}} \bar{h}_{Q_i}^{(v)}\,
(-i\overleftarrow{\!\not\!\! D})\,
\Gamma h_{Q_i}^{(v)} +{\cal O}(\lambda^2) \, ,
\label{JCOMB}
\end{equation}  
where the covariant derivative $\overleftarrow{\!\not\!\! D}$ acts on the
field on the left. The expressions in eqs.~(\ref{QCOMB}), (\ref{LHCOMB}) and
(\ref{JCOMB}) are identical in form to the analogous quantities in  HQET. 
However, there is one important difference: these expressions represent the 
combined expansions in powers of $\lambda$ and not the $1/m_Q$ expansion of 
HQET. In other words, by defining fields $h_{Q}^{(v)}$ and $H_{Q}^{(v)}$ 
appropriately, we were able to re-sum implicitly all the $m_N/m_Q$ corrections.

A typical magnitude of the momentum, $k$, carried by the effective field 
$h_{Q}^{(v)}$ is of order $\bar{\Lambda}\equiv m_{\Lambda_Q}-(m_Q+m_N) 
\sim \lambda^0$. As a result, operators containing $n$ powers of the covariant
derivative $D$ are of order $(k/m_Q)^n \sim \lambda^n$ in the combined limit.
For example, the second and third terms in eq.~(\ref{JCOMB}) are of relative 
order $\lambda$, so that they contribute only at NNLO in the combined 
expansion. Therefore, only one operator---$\bar{h}_{Q_j}^{(v)}
\Gamma h_{Q_i}^{(v)}$---contributes at LO and NLO.
 
The effective vector and axial currents at leading and next-to-leading order 
are given by:
\begin{eqnarray}
\bar{c} \gamma^\mu b \,\, (y=0) & = &
\bar{h}_{c}^{(v)}\gamma^\mu h_{b}^{(v)} \,\, (y=0) + 
{\cal O}(\lambda) \, ,
\nonumber \\
\bar c \gamma^\mu \gamma_5 b \,\, (y=0) & = &
\bar{h}_{c}^{(v)}\gamma^\mu \gamma_5 h_{b}^{(v)} \,\, (y=0) + 
{\cal O}(\lambda) \, .
\label{VAEFF}
\end{eqnarray}
The heavy quark effective currents, eq.~(\ref{VAEFF}), have the same form at 
$y=0$ regardless of whether the heavy quark effective fields 
${h}_{c}^{(v^\prime)}$ and $h_{b}^{(v)}$ are defined at the same 4-velocity, 
$v^\prime=v$, or at two different values, $v^\prime \neq v$.

Now it is easy to show the self-consistent scaling rules of the 
electroweak matrix elements, eq.~(\ref{CRLHS}). The effective operators
corresponding to the time component of the vector current and spatial
components of the axial current scale as $\lambda^0$:
\begin{eqnarray}
h_{c}^{\dag (v)} \, h_{b}^{(v)} & \sim & \lambda^0 \, , 
\nonumber \\
h_{c}^{\dag (v)} \, \Sigma^i \, h_{b}^{(v)} & \sim & \lambda^0 \, ,
\label{V0AVscaling}
\end{eqnarray}
where the scaling of the axial operators follow because 
$\Sigma^i = \gamma^0 \gamma^i \gamma_5$ act only
on the upper components of the effective heavy quark fields which are of order
$\lambda$. On the other hand, spatial components of the vector current and
time component of the axial current are of order $\lambda^{3/4}$:
\begin{eqnarray}
\bar{h}_{c}^{(v)} \,\gamma^i h_{b}^{(v)} =  
h_{c}^{\dag (v)} \,\alpha^i \, h_{b}^{(v)} & \sim & \lambda^{3/4} \, , 
\nonumber \\
\bar{h}_{c}^{(v)} \,\gamma^0 \gamma_5 \, h_{b}^{(v)} =
h_{c}^{\dag (v)} \,\gamma_5 \, h_{b}^{(v)} & \sim & \lambda^{3/4} \, 
\label{VVA0scaling}
\end{eqnarray}
since the matrices $\alpha^i$ and $\gamma_5$ mix the upper and lower components
of the effective heavy quark fields which brings an additional power
of $\lambda^{3/4}$. 

As shown in Sec.~\ref{SD}, the counting rules in eqs.~(\ref{V0AVscaling}) and
(\ref{VVA0scaling}) lead to the self-consistent scaling rules for the 
semileptonic form factors, eqs.~(\ref{FFCR}), (\ref{KLCR}), (\ref{NCR}) and
(\ref{MCR}).  As a result, the dominant form factors at LO and NLO for the
$\Lambda_b \to \Lambda_c \ell \bar{\nu}$ decay are $F_1$ and $G_1$  given by,
\begin{eqnarray}
\langle \Lambda_c (\vec{v}^\prime)|\bar{c}\gamma^0 b|\Lambda_b (\vec{v}) 
\rangle =
\langle \Lambda_c (\vec{v}^\prime)|h_{c}^{\dag (v)} \, h_{b}^{(v)}
|\Lambda_b (\vec{v}) \rangle
&=& F_1 u^{\dag}_c (\vec{v}^\prime) u_b (\vec{v}) 
\left (1 + {\cal O}(\lambda) \right ) \, ,
\nonumber \\
\langle \Lambda_c (\vec{v}^\prime)|\bar{c}\gamma^i \gamma_5 b|\Lambda_b 
(\vec{v}) \rangle =
\langle \Lambda_c (\vec{v}^\prime)|h_{c}^{\dag (v)} \, \Sigma^i \, h_{b}^{(v)}
 |\Lambda_b (\vec{v}) \rangle &=&
G_1 u^{\dag}_c (\vec{v}^\prime) \Sigma^i u_b (\vec{v}) 
\left (1 + {\cal O}(\lambda) \right ) \, ,
\label{F1G1EFF} 
\end{eqnarray} 
where the effective heavy quark operators, eq.~(\ref{VAEFF}), were used.
The effective operator $h_{c}^{\dag (v)} \, \Sigma^i \, h_{b}^{(v)}$ acts on 
the heavy quark spin degrees of freedom due to the $\Sigma^i$ matrix. As
discussed in Sec.~\ref{S}, the heavy quark spin decouples from the dynamics
of the collective degrees of freedom---the motion of the brown muck relative
to the heavy quark---at LO and NLO in the combined limit. Hence, up to NNLO
the effect of the heavy quark effective operator
$h_{c}^{\dag (v)} \, \Sigma^i \, h_{b}^{(v)}$ in the Hilbert space of the
low-lying heavy baryon states is identical to that of the effective 
operator $h_{c}^{\dag (v)} \,h_{b}^{(v)}$. As a result, the dominant
vector and axial form factors---$F_1$ and $G_1$---are equal up to NNLO in the
combined limit.

Similarly, the pair of the form factors $K_1$ and $L_1$ (dominant in the 
$\Lambda_{b} \to \Lambda^{*}_{c} \ell \bar{\nu}$ decay), and $N_1$ and $M_1$
(dominant in the $\Lambda_{b} \to \Lambda^{*}_{c1} \ell \bar{\nu}$ decay) are 
equal up NNLO in the combined limit.

\end{document}